\pdfoutput=1
\documentclass[mathleft,fleqn,%
]{an}
\usepackage{graphicx}
\usepackage{url}
\usepackage[varg]{txfonts}
\usepackage{natbib}
\bibpunct{(}{)}{;}{a}{}{,}
\newcommand{\kms}{\mbox{km\,s$^{-1}$}}

\begin{document}

\Pagespan{1}{}
\Yearpublication{2014}%
\Yearsubmission{2014}%
\Month{0}%
\Volume{999}%
\Issue{0}%
\DOI{asna.201400000}%

\title{Spectroscopic and photometric confirmation of chromospheric activity in four stars}

\author{O.\,\"Ozdarcan\inst{1}\fnmsep\thanks{Corresponding author:
        {orkun.ozdarcan@ege.edu.tr}}
\and  H.\,A. Dal\inst{1}
}
\titlerunning{Chromospheric activity of four cool stars}
\authorrunning{O.\,\"Ozdarcan \& H.\,A. Dal}
\institute{
Ege University, Science Faculty, Department of Astronomy and Space Sciences, 
35100 Bornova, \.{I}zmir, Turkey.}

\received{XXXX}
\accepted{XXXX}
\publonline{XXXX}


\abstract{ We present analysis of medium resolution optical spectra and long term $V$ 
band photometry of four cool stars, \object{BD+13 5000}, \object{BD+11 3024}, 
\object{TYC\,3557-919-1} and \object{TYC\,5163-1764-1}. Our spectroscopic analysis
reveals that the stars are giant or sub-giant from K0 or K1 spectral type, and all of them
exhibit emission features in their Ca\,{\sc ii} H\& K lines. These features appear to be modulated
with the rotation of the stars. Except \object{BD+11 3024}, we observe that the radial velocities 
of the target stars are not stable, which suggests that each of them might be a member of a binary 
system. Global analysis of photometric data indicates clear cyclic variation for \object{BD+13\,5000} 
and \object{TYC\,5163-1764-1} with a period of 8.0$\pm$0.3 and 5.04$\pm$0.04 year, respectively. 
Besides that, we observe a dramatic increase ($\sim$0\fm7) in the mean brightness of \object{BD+11 3024},
accompanied with a $2.87\pm0.12$ cyclic variation, embedded into the global brightening trend, which
indicates possible multiple cycles on this star.}

\maketitle

\section{Introduction}\label{S1}

Advent of the Automatic Photoelectric Telescopes \citep[APT, see, e.g.][]{APTs_Henry_1995ASPC, APTs_Strassmeier_1997PASP} 
enabled precise and efficient long term photometric observations
of chromospherically active stars, which exhibit emission in their Ca\,{\sc ii} H\& K lines.
Observed stars are usually brighter than 10 magnitude in $V$ band. Nowadays, these stars have 
continuous photometry for more than twenty years, which were used in many studies on analysing
photometric activity cycles \citep{Olah_cycles_2009A&A, Jetsu_Henry_long_term_data_2017ApJ}, 
indirect surface imaging \citep{Roettenbacher_LC_inversion_2011AJ}, and relation between 
light curve properties and photometric periods \citep[see, e.g.][]{FG_IS_Fekel_et_al_2002AJ, 
Ozdarcan_V2075Cyg_2010AN, V1149Ori_Fekel_et_al_2005AJ}.

Further contribution in terms of collecting continuous photometry came from photometric surveys,
such as The All Sky Automated Survey \citep[ASAS,][]{ASAS_Pojmanski_1997, ASAS_Pojmanski_2002AcA, 
ASAS3_Pojmanski_2005AcA}, and Northern Sky Variability Survey \citep[NSVS,][]{NSVS_Wozniak_2004AJ}. 
From these databases, numerous variable stars, which are fainter relative to the stars observed by 
APTs, were discovered. \object{BD+11 3024}, \object{BD+13 5000}, \object{TYC\,3557-919-1} and 
\object{TYC\,5163-1764-1} are such targets.

BD+11\,3024 exhibits strong X-ray emission \citep{Zickgraf_et_al_2003A&A} and photometric 
variability with a period of 21\fd69 \citep{gsc3557_gsc969_Bernhard_2008OEJV}. 
\object{TYC\,3557-919-1} was listed in the catalogue of \citet{Haakonsen_et_Al_2009ApJS}, where the 
near-infrared counter parts of X-ray sources in ROSAT Bright Source Catalogue 
\citep{ROSAT_Voges_et_al_1999A&A} were presented. Photometric variability of the star was reported 
by \citet{gsc3557_gsc969_Bernhard_2008OEJV} with a period of 25\fd08. 
BD+13\,5000 took place in the catalogue published by \citet{Boyle_et_al_1997MNRAS}, which provides 
optical counterpart of some X-ray sources in the ROSAT catalogue. \citet{Hoffman_et_Al_2009AJ} 
identified the star as a new variable. \citet{Kiraga_ASAS_ROSAT_sources_2012AcA} identified it
as a rotating variable with a period of 18\fd14, while \citet{gsc5163_gsc1159_2011OEJV} classified
the star as chromospherically active. Among our target stars in this work, \object{TYC\,5163-1764-1}
has the least literature information. The star was identified as an X-ray source in ROSAT catalogue
\citep{ROSAT_Voges_et_al_1999A&A}, and then its photometric variability was reported by
\citet{gsc5163_gsc1159_2011OEJV} with a variation period of 26\fd08. All the four targets
have no detailed study on their atmospheric properties and photometric characteristics.

In the scope of this study, we obtained medium resolution optical spectra of these targets to 
determine their spectral characteristics and investigate spectroscopic binarity of them. 
Furthermore, we collected differential $V$ band photometry of the targets from three different 
sources. Collected data enabled us to investigate seasonal and long term photometric behaviour 
of the targets, as well as their seasonal photometric periods. In the next section, we describe our
spectroscopic observations, data reductions and analysis. Sources of collected photometric data,
together with global and seasonal photometric analysis are given in Section~\ref{S3}. In the last
section, we summarize our findings on each target.

\section{Spectroscopy}\label{S2}

\subsection{Observations and data reduction}\label{S2.1}

We carried out optical spectroscopic observations of the program stars with Turkish Faint Object 
Spectrograph Camera (TFOSC\footnote{\url{http://www.tug.tubitak.gov.tr/rtt150_tfosc.php}}) 
attached to the 1.5 m Russian -- Turkish telescope at T\"UB\.ITAK National Observatory (TNO). Back illuminated
2048 $\times$ 2048 pixels CCD camera with a pixel size of 15 $\times$ 15 $\mu m^{2}$ was used together
with the spectrograph. Using \'echelle mode of TFOSC, we were able to record spectra between 
3900 -- 9100 \AA\@ in 11 \'echelle orders, which provided spectral resolution of 
R = $\lambda/\Delta\lambda$ $\sim$ 2800 around 6500 \AA\@.

We follow standard procedure for reducing \'echelle spectra, which basically includes bias correction,
flat-field division of Fe-Ar calibration frames and science frames, followed by scattered light correction
and cosmic rays removal, and finally extraction of spectra from \'echelle orders. Wavelength calibration
of reduced and extracted science spectra are done via Fe-Ar images, and wavelength calibrated science spectra 
are normalized to the unity by using 4th or 5th order cubic spline function.


\subsection{Radial velocities}\label{S2.2}

We give a brief log of spectroscopic observations in Table~\ref{T1}. Note that, in addition to the target star
observations, we obtained optical spectra of \object{35\,Peg} (K0 III) and \object{$\theta$\,Psc} (K1 III) with
the same instrumental set-up, and used them as spectroscopic comparison and radial velocity template.
We measure radial velocities
of the targets by cross-correlating each of their spectrum with the observed templates using the technique
described in \citet{fxcor_Tonry_Davis_1979}. We apply the technique by using $fxcor$ task 
\citep{fxcor_Fitzpatrick_1993ASPC} of IRAF\footnote{The Image Reduction and Analysis Facility is hosted by the National 
Optical Astronomy Observatories in Tucson, Arizona at URL iraf.noao.edu.} software. We use \object{$\theta$\,Psc} as radial velocity
template for \object{BD+13\,5000}, while \object{35\,Peg} is the template for the remaining target
stars. Resulting cross-correlation functions exhibit sharp and single peak for each target, indicating
either these stars are single, or SB1 systems which can not be resolved in our resolution.

\begin{table}
\caption{Log of spectroscopic observations. Exposure time is in second, signal-to-noise ratio 
(S/N) is measured around 5500 \AA\@. Radial velocity (V$_{r}$) and its standard error 
($\sigma_{V_{r}}$) are given in \kms\@.}\label{T1}
\scriptsize
\begin{center}
\begin{tabular}{cccccc}
\hline\noalign{\smallskip}
Star  &  HJD  & Exp.  & S/N &  V$_{r}$ & $\sigma_{V_{r}}$ \\
      &  (24 57000+)  & (s)  &   &    &   \\
\hline\noalign{\smallskip}
                        &    378.2194  &  3600 &   208  &  $-$39.3  &  4.1  \\
                        &    604.5612  &  3600 &   138  &  $-$43.9  &  2.4  \\
\object{BD+13\,5000}    &    605.4999  &  3600 &   177  &  $-$37.3  &  5.8  \\
                        &    671.4607  &  3600 &   239  &     32.9  &  2.9  \\
                        &    672.4661  &  3600 &   225  &     46.3  &  1.6  \\
\hline\noalign{\smallskip}
                        &    604.4006  &  3600 &   107  &  $-$38.8  &  2.0  \\
                        &    605.4410  &  3600 &   122  &  $-$36.1  &  2.2  \\
                        &    671.4051  &  3600 &   182  &      9.8  &  1.7  \\
\object{TYC\,3557-919-1}&    672.4114  &  3600 &   166  &     12.9  &  2.2  \\
                        &    853.5443  &  1095 &    98  &  $-$24.4  &  2.2  \\
                        &    854.4724  &  3200 &    70  &  $-$27.3  &  3.4  \\
\hline\noalign{\smallskip}
                            & 604.4871  &  3600  &  107  &     36.1  &   4.2  \\
\object{TYC\,5163-1764-1}   & 605.3279  &  3600  &  41   &     27.4  &   4.3  \\
                            & 671.3019  &  3600  &  165  &  $-$13.2  &   3.3  \\
                            & 672.2491  &  3600  &  179  &  $-$12.2  &   3.5  \\
\hline\noalign{\smallskip}
                       &    604.3434   &  3600  &  200  &    11.1  &  3.9  \\
\object{BD+11\,3024}   &    605.3820   &  3600  &  200  &     8.9  &  4.6  \\
                       &    832.6153   &  3600  &  248  &     3.9  &  2.7  \\
                       &    854.4324   &  2700  &  276  &  $-$6.3  &  4.3  \\
\noalign{\smallskip}\hline
\end{tabular}
\end{center}
\end{table}

\begin{figure*}[!htb]
\centering
{\includegraphics[angle=0,scale=0.50,clip=true]{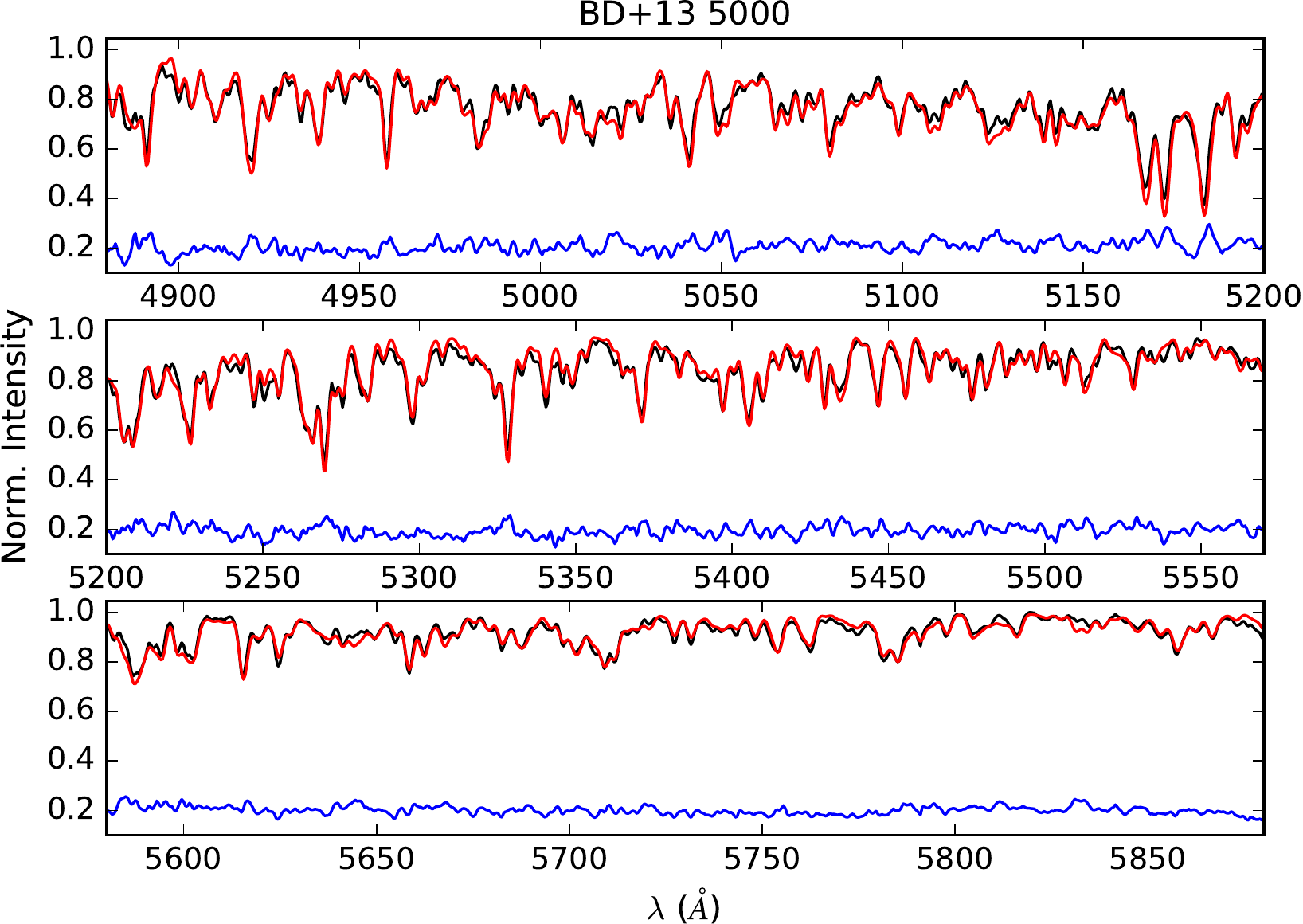}}
{\includegraphics[angle=0,scale=0.50,clip=true]{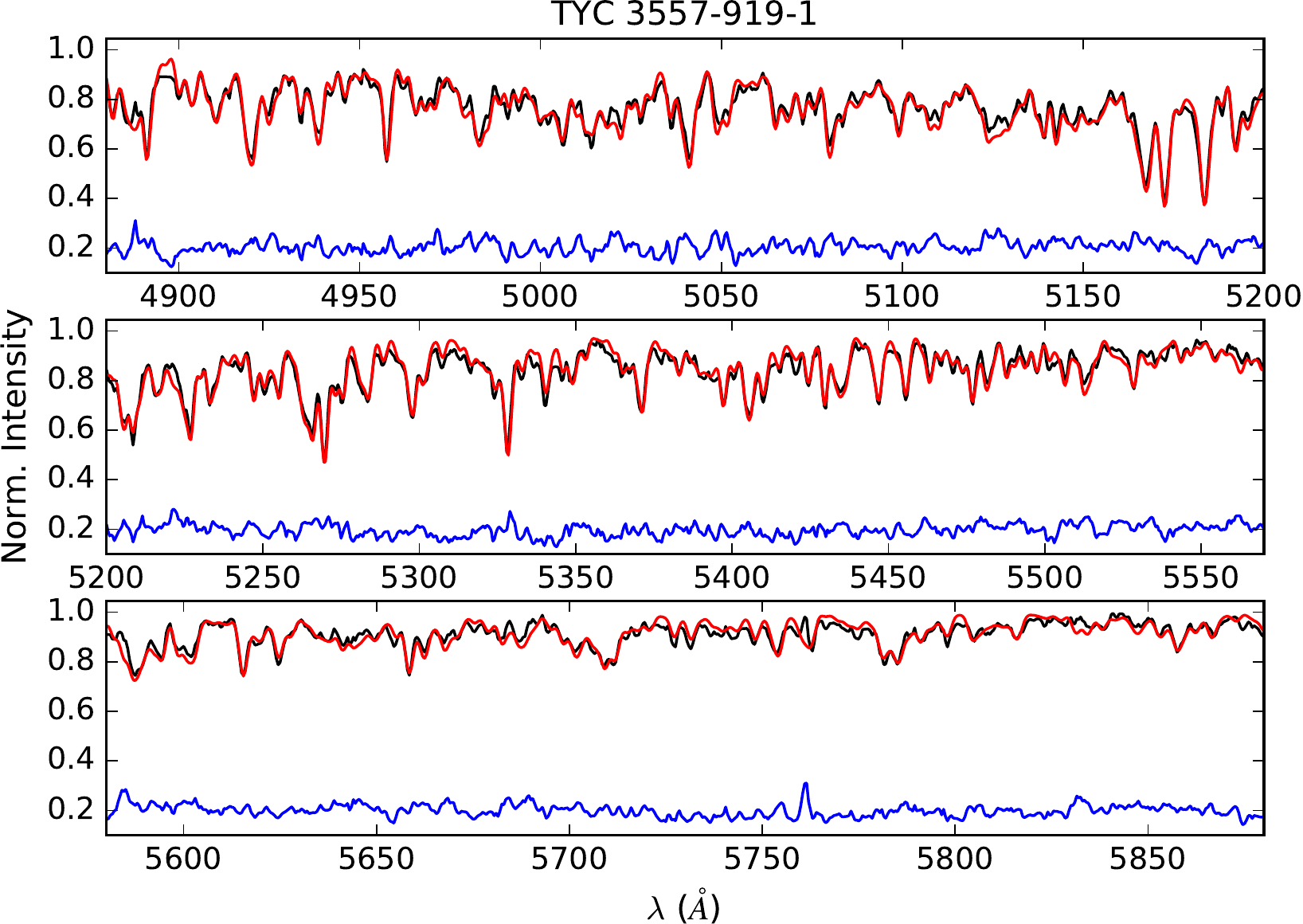}}\vspace{0.3cm} \\
{\includegraphics[angle=0,scale=0.50,clip=true]{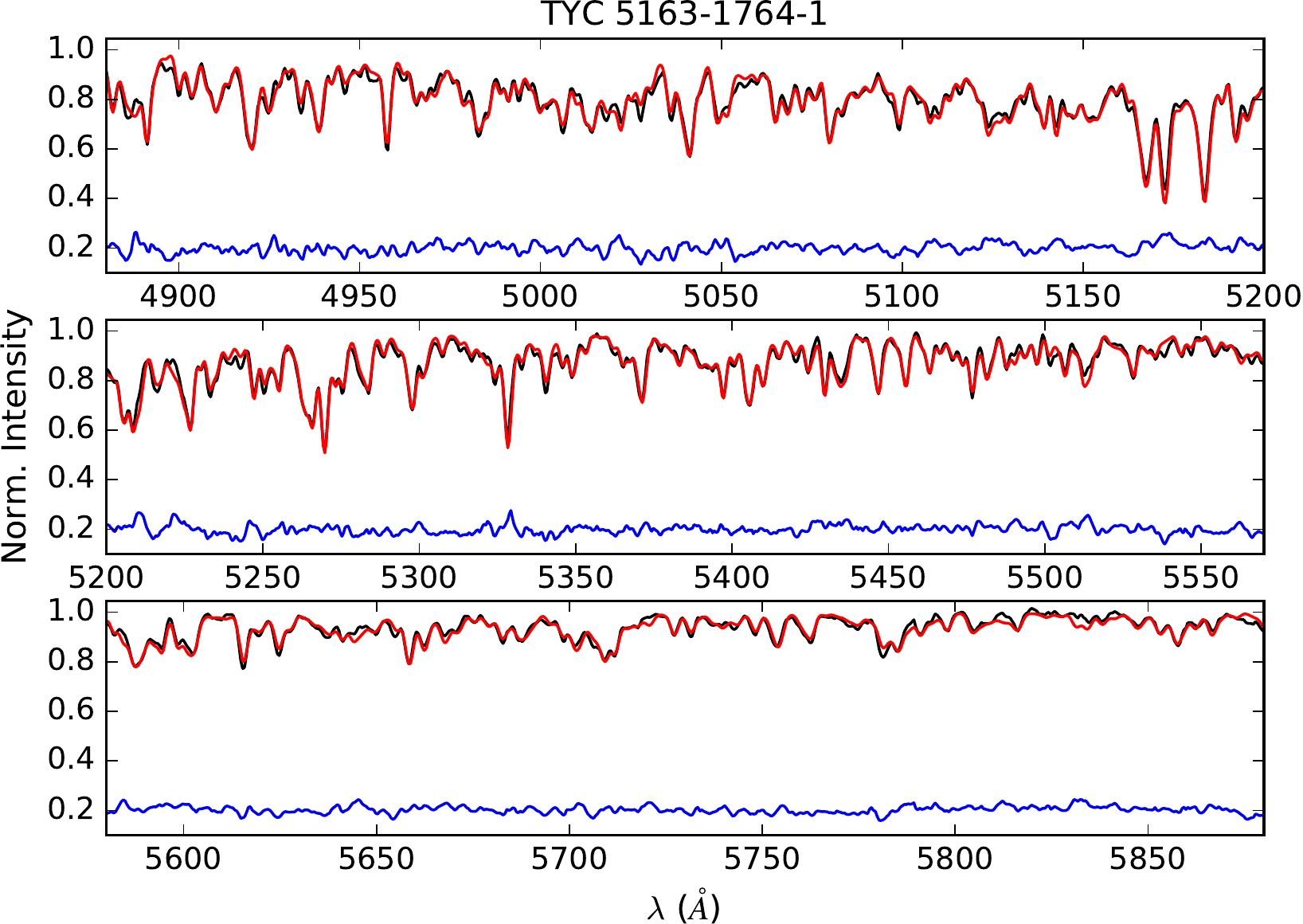}}
{\includegraphics[angle=0,scale=0.50,clip=true]{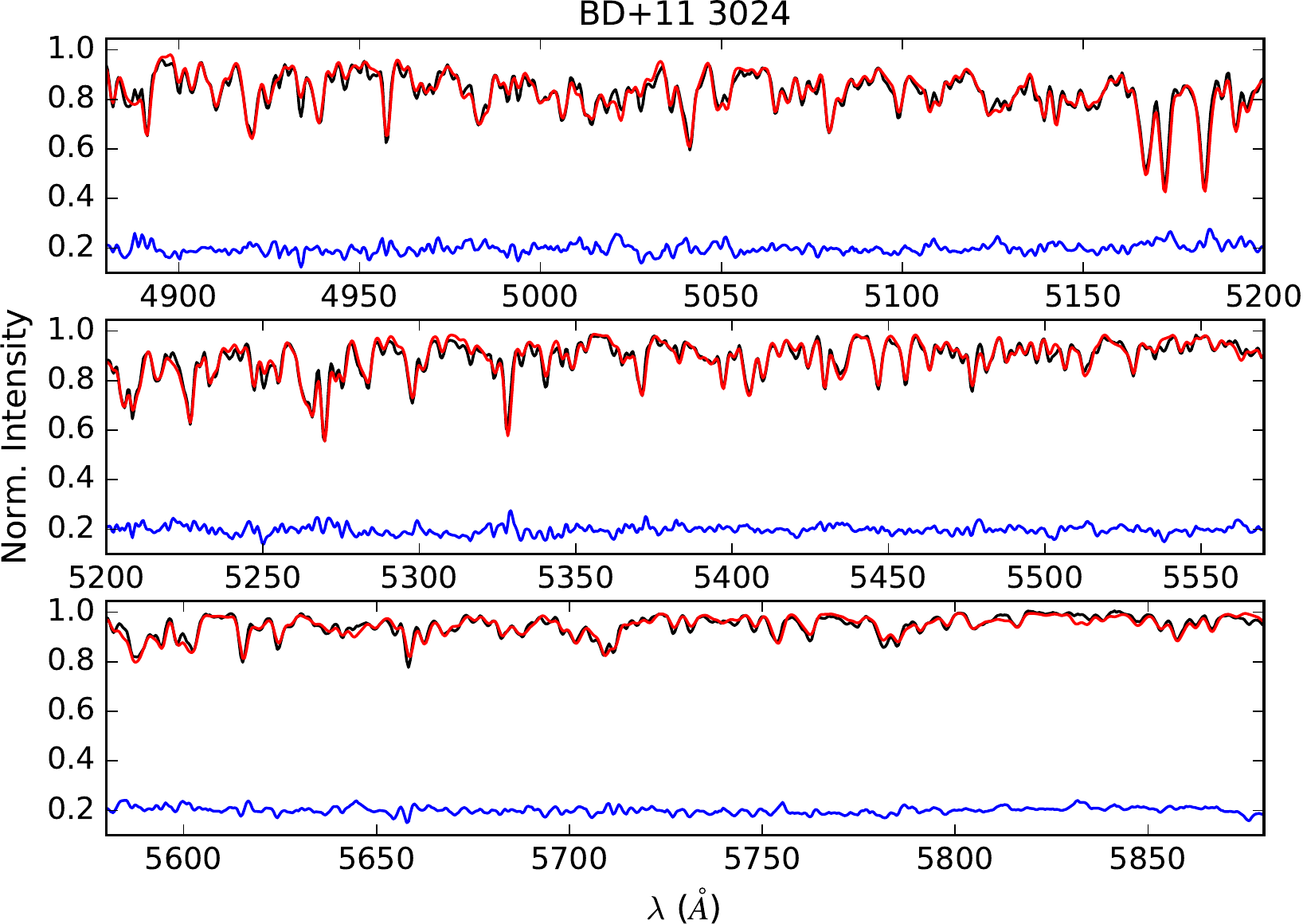}}
\caption{Observed (black) and best-matched synthetic spectrum (red) for each target stars.
Residuals from the best-matched spectrum are shown in blue and shifted by 0.2 upwards for 
the sake of simplicity.}\label{F1}
\end{figure*}

Inspecting Table~\ref{T1}, one can notice non-stable radial velocities for each target. Difference
between measured maximum and minimum radial velocities is the smallest and about 17 \kms\@ for
\object{BD+11\,3024}, but much larger for the others. This could be result of a motion
on a Kepler orbit, i.e. each star may actually be a member of an SB1 system. Although the clear
difference between measured radial velocities of each individual star, we can not check this
possibility with a preliminary orbital solution due to the less number of observed spectra sparsely 
distributed in time. However, we plot phase-folded radial velocities of each target in the appendix,
for visual evaluation of radial velocity patterns with respect to the calculated photometric periods
(see Section~\ref{S3.3.2} and Table~\ref{T5}).

\subsection{Spectral types and features}\label{S2.3}

We compare spectra of target stars with the observed template spectra of \object{35\,Peg} and \object{$\theta$\,Psc} 
to determine spectral type of each target. We choose a single observed spectrum for each target,
which has the highest S/N, for comparison. In the first step we observe that the spectrum of 
\object{35\,Peg} provides close match to the spectra of \object{TYC\,3557-919-1}, 
\object{TYC\,5163-1764-1} and \object{BD+11\,3024}, thus indicating K0 III spectral type, while 
spectrum of \object{$\theta$\,Psc} is similar to the spectra of \object{BD+13\,5000} and pointing out 
K1 III spectral type. In the second step, we adopt previously determined atmospheric parameters of
\object{35\,Peg} and \object{$\theta$\,Psc} \citep{35Peg_thetaPsc_Sharma2016}, as initial atmospheric
parameters of corresponding target star, and switch to the spectrum synthesizing method to find the 
best matching synthetic spectrum to the observed spectrum of each target. Atmospheric parameters of the
best matching synthetic spectrum is adopted as the final atmospheric parameters of the corresponding
target.

We use the latest version of python framework $iSpec$ \citep{iSpec_Cuaresma_2014A&A}, which includes
different spectrum synthesis codes, model atmospheres and line lists. In our case, we adopt 
SPECTRUM\footnote{http://www.appstate.edu/$\sim$grayro/spectrum/spectrum.html} code 
\citep{spectrum_gray_1994} in conjunction with MARCS model atmospheres \citep{MARCS_Gustafsson_2008}
and the updated line list from Vienna Atomic Line Database \citep[VALD, ][]{VALD3_Ryabchikova_2015}.
We calculate synthetic spectra for effective temperatures between 4000--5500 K with a step of 250 K,
log$g$ value between 2.0--4.5 with a step of 0.5, and metallicity values ([M/H]) between $-$1.0 and 
0.0 in steps of 0.25. We convolve each synthesized spectrum with a proper Gaussian line-spread function to 
match the resolution of TFOSC spectra. Then, we re-normalize the observed spectrum with respect to 
the instrumentally broadened synthetic spectrum, in order to compensate continuum level uncertainty coming 
from normalization by cubic splines in IRAF. Finally, we compare synthesized spectrum with the re-normalized 
observed spectrum by considering residuals. Since numerical fitting methods are not very efficient due 
to the resolution of TFOSC spectra, we only calculate synthetic spectra for available MARCS models, and
refrain from any interpolation between the available model atmospheres. In the whole process, we keep 
micro-turbulence velocity fixed at 2 \kms\@. Considering resolution and S/N of TFOSC spectra, and grid 
steps of MARCS models for effective temperature, log$g$ and metallicity, we estimated uncertainties as 200 K in 
$T_{\rm eff}$, 0.5 in log$g$ and 0.25 in $[$M/H$]$. We list the atmospheric parameters of the synthetic
spectra, that provides the closest match to the observed spectra of the target stars, in Table~\ref{T2}.
In Figure~\ref{F1}, we plot observed and best-matched spectra at optical wavelengths, together with 
residuals from the best-matched synthetic spectra.

\begin{table}
\caption{Final atmospheric parameters of target stars. Sp denotes estimated spectral type.}\label{T2}
\small
\begin{center}
\begin{tabular}{lcccl}
\hline\noalign{\smallskip}
Star &  $T_{\rm eff}$ & log$g$ & $[$M/H$]$ & Sp. \\
     &        (K)     &  (cgs) &           &     \\
\hline\noalign{\smallskip}
\object{BD+13\,5000}       & 4750 & 3.5 &  0.0 & K0IV \\
\object{TYC\,3557-919-1}   & 4750 & 3.0 &  0.0 & K0III-IV \\
\object{TYC\,5163-1764-1}  & 4500 & 2.5 & -0.5 & K2III \\
\object{BD+11\,3024}       & 4750 & 3.0 & -0.5 & K0III-IV \\
\noalign{\smallskip}\hline
\end{tabular}
\end{center}
\end{table}

\begin{figure*}[!htb]
\centering
{\includegraphics[angle=0,scale=0.45,clip=true]{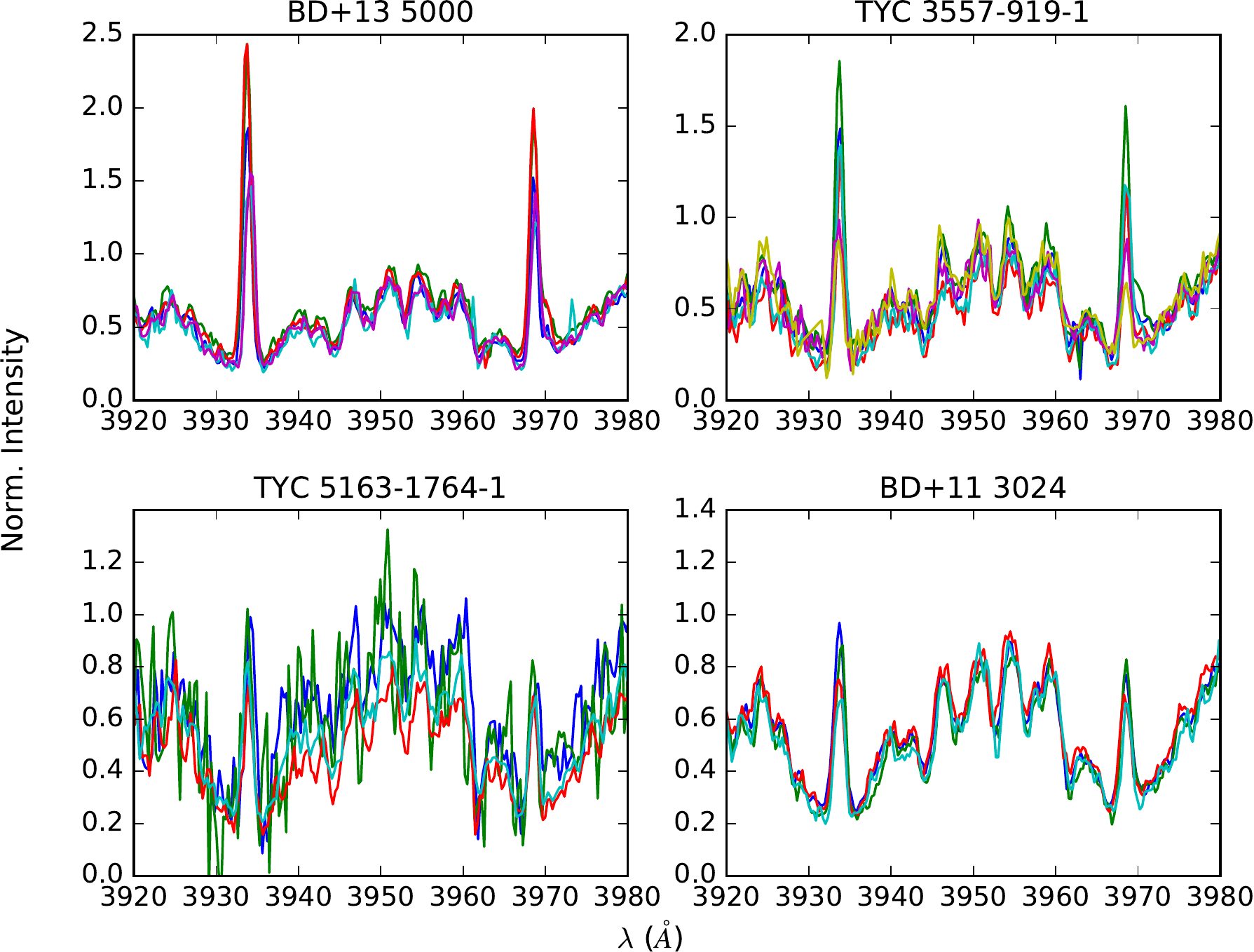}}\hspace{0.2cm}
{\includegraphics[angle=0,scale=0.45,clip=true]{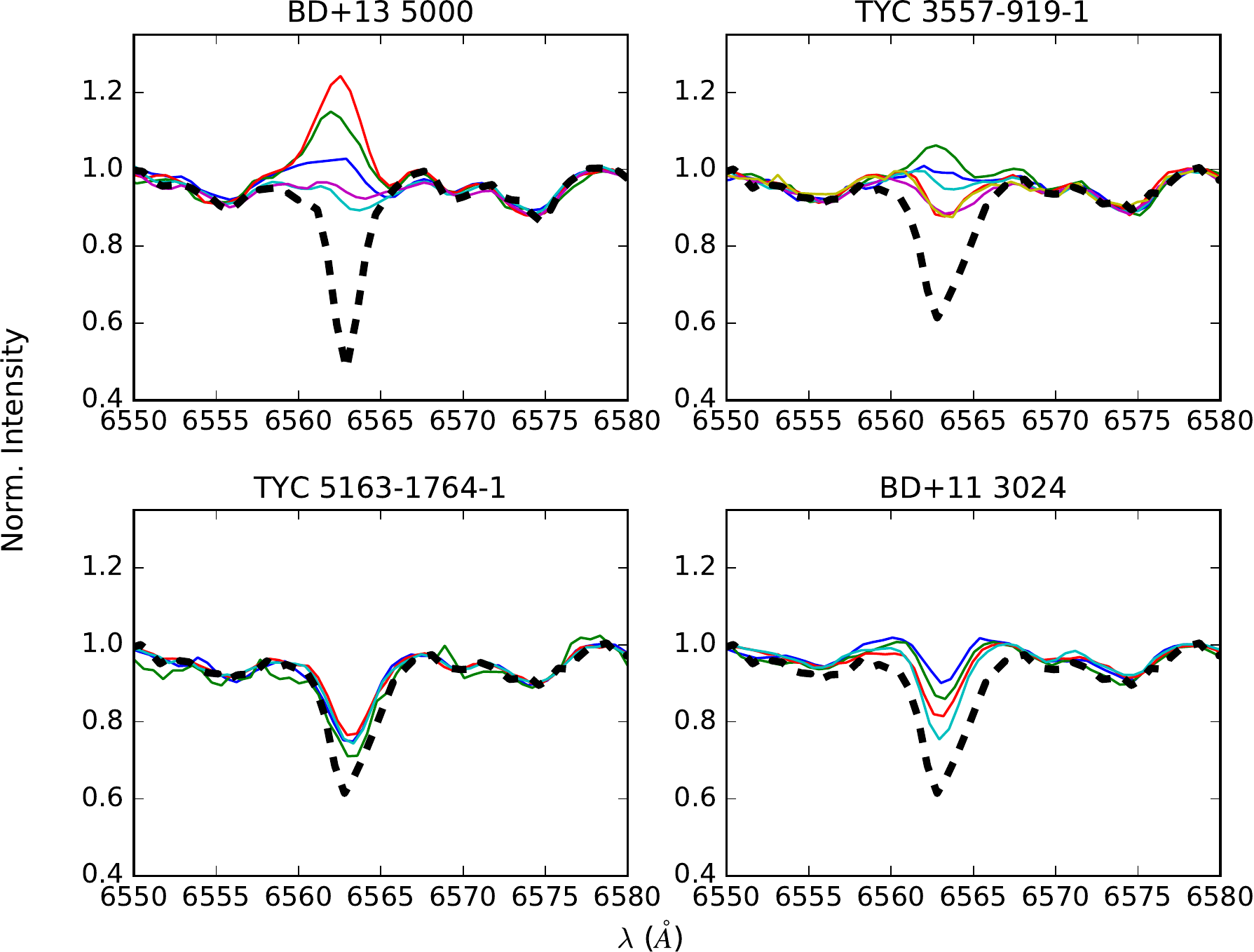}}
\caption{Observed spectra around Ca\,{\sc ii} H\& K lines (left four plots) and H$_{\alpha}$
line (right four plots). For each star, we plot all observed spectra with continuous lines in 
different colors, indicating different Julian dates given in Table~\ref{T1}. In H$_{\alpha}$
plots, note that we over plot observed spectrum of \object{35\,Peg} with thick dashed line for
\object{BD+13\,5000}, while the same thick dashed line denotes observed spectrum of \object{$\theta$\,Psc}
in the remaining plots.}\label{F2}
\end{figure*}

We observe noticeable spectral characteristics of H$_{\alpha}$ and Ca\,{\sc ii} H\& K lines, which
are sensitive to the chromospheric activity in cool stars. In Figure~\ref{F2}, we plot observed 
spectra around these lines. For all stars, Ca\,{\sc ii} H\& K lines are in form of emission. 
Very strong emission in Ca\,{\sc ii} H\& K lines, which exceeds continuum, and varying H$_{\alpha}$
profiles are remarkable for \object{BD+13\,5000} and \object{TYC\,3557-919-1}, while 
\object{BD+11\,3024} and \object{TYC\,5163-1764-1} show variable H$_{\alpha}$ line profiles, 
but usually in form of filled absorption, which can be concluded by comparing their line profiles 
with the template star spectrum. Considering variable strength of emission features at different 
observation epochs, these features appear to be modulated with the rotation of the star. These 
spectral features and atmospheric parameters of the target stars clearly indicate chromospheric 
activity.

\section{Photometry}\label{S3}

We collect differential $V$ photometry of our target stars with respect to a proper nearby 
comparison and check stars. We list basic data of all program stars in Table~\ref{T3}.

\begin{table*}
\caption{Basic data of program stars. $V$ magnitudes and $B-V$ colours are from Tycho-2 catalogue
\citep{Tycho_2_cat_Hog_et_al2000A&A}, while $JHK$ magnitudes are from 2MASS All Sky Catalog of 
point sources \citep{2MASS_Cutri_et_al_2003yCat}.}\label{T3}
\begin{center}
\begin{tabular}{ccccccccc}
\hline\noalign{\smallskip}
Star  &  Identifier  & $\alpha$ (2000) & $\delta$ (2000) &  $V$  &  $B-V$ & $J$ & $H$ & $K$ \\
\hline\noalign{\smallskip}
Variable & \object{BD+13\,5000}  & 22$^{h}$~50$^{m}$~24$^{s}$ &  $+$14$^{\circ}$~31$'$~43$''$ & 10\fm61 & 1\fm13 & 8\fm455 & 7\fm895 & 7\fm748 \\
Comparison & \object{2MASS\,22502241+1434264} & 22$^{h}$~50$^{m}$~22$^{s}$ &  $+$14$^{\circ}$~34$'$~26$''$ & --- & --- & 11\fm105 & 10\fm946 & 10\fm858 \\
Check     & \object{BD+13\,5001} & 22$^{h}$~50$^{m}$~30$^{s}$ &  $+$14$^{\circ}$~36$'$~45$''$ &  9\fm94 & 0\fm31 & 9\fm276 & 9\fm183 & 9\fm124 \\

  &  &  &  &   &  &  &  &  \\

Variable & \object{TYC\,3557-919-1}  &  19$^{h}$~43$^{m}$~40$^{s}$ & $+$46$^{\circ}$~40$'$~03$''$ & 11\fm22 & 1\fm00 & 9\fm232 & 8\fm684 & 8\fm536 \\
Comparison & \object{2MASS\,19432561+4641206}  &  19$^{h}$~43$^{m}$~26$^{s}$ & $+$46$^{\circ}$~41$'$~21$''$ & --- & --- & 10\fm135 & 9\fm730 & 9\fm617 \\
Check     & \object{2MASS\,19430979+4642507}  &  19$^{h}$~43$^{m}$~10$^{s}$ & $+$46$^{\circ}$~42$'$~51$''$ & --- & --- & 11\fm947 & 11\fm852 & 11\fm827 \\

  &  &  &  &   &  &  &  &  \\

Variable & \object{TYC\,5163-1764-1} & 20$^{h}$~23$^{m}$~34$^{s}$ & $-$01$^{\circ}$~45$'$~02$''$ & 11\fm335  & 1\fm33  &  --- & --- & --- \\
Comparison & \object{TYC\,5163-1960-1}  &  20$^{h}$~23$^{m}$~06$^{s}$ & $-$01$^{\circ}$~35$'$~08$''$ & 11\fm40 & 1\fm41 & 9\fm625 & 9\fm101 & 8\fm984 \\
Check     &  \object{TYC\,5163-1652-1}  &  20$^{h}$~23$^{m}$~09$^{s}$ & $-$01$^{\circ}$~42$'$~16$''$ & 11\fm70 & 1\fm06 & 9\fm924 & 9\fm424 & 9\fm29 \\

  &  &  &  &   &  &  &  &  \\

Variable & \object{BD+11\,3024} & 16$^{h}$~41$^{m}$~53$^{s}$ & $+$11$^{\circ}$~40$'$~21$''$ & 10\fm13 & 1\fm10 & 7\fm890 & 7\fm294 & 7\fm077 \\
Comparison & \object{BD+11\,3026p} & 16$^{h}$~42$^{m}$~26$^{s}$ & $+$11$^{\circ}$~40$'$~55$''$ & 10\fm59 & 0\fm56 & 9\fm467 & 9\fm188 & 9\fm132 \\
Check     & \object{BD+11\,3026} & 16$^{h}$~42$^{m}$~28$^{s}$ & $+$11$^{\circ}$~41$'$~35$''$ & 9\fm90 & 0\fm44 & 8\fm953 & 8\fm773 & 8\fm698 \\

\noalign{\smallskip}\hline
\end{tabular}
\end{center}
\end{table*}

Differential $V$ photometry comes from three sources. The first and the second sources are 
The All Sky Automated Survey (ASAS), and All-Sky Automated Survey for Supernovae Sky Patrol 
\citep[ASAS-SN,][]{ASAS_SN_2014ApJ, ASAS_SN_2017arXiv170607060K} databases, and 
the third source is our Bessell $V$ CCD observations obtained at TNO.

\subsection{ASAS and ASAS-SN photometry}\label{S3.1}

We extract all available $V$-band photometric data of program stars from ASAS and ASAS-SN databases. 
Since we only have differential magnitudes from T60 observations, we convert standard magnitudes to 
differential magnitudes in order to evaluate observations on a common scale. We calculate average 
standard $V$ magnitude of each comparison star in the databases, and subtract it from each individual 
standard $V$ magnitudes of corresponding target star, thus obtain differential magnitudes in the 
sense of variable-minus-comparison. We estimate the precision of differential magnitudes with respect 
to the check-minus-comparison magnitudes. For check-minus-comparison measurements, we consider 
observing nights where simultaneous measurements from both check and comparison stars are available 
at exactly the same Julian date, and then obtain check-minus-comparison magnitudes. We list a brief 
information on collected data from ASAS and ASAS-SN databases in Table~\ref{T4}.

\begin{table}
\caption{Brief information on data collected from ASAS, ASAS-SN databases and T60
telescope. N and $\sigma$ show number of data points and standard deviation calculated 
from check-minus-comparison measurements for corresponding star.}\label{T4}
\scriptsize
\begin{center}
\begin{tabular}{cccc}
\hline\noalign{\smallskip}
     &  \multicolumn{3}{c}{ASAS}   \\
Star &  HJD range & N & $\sigma$ \\
     & (24 00000+)&   &  (mag)   \\
\hline\noalign{\smallskip}
\object{BD+13\,5000}      & 52765-55167 & 252 & 0.066 \\
\object{TYC\,3557-919-1}  & --- & --- & ---  \\
\object{TYC\,5163-1764-1} & 51985-55145 & 369 & 0.045 \\
\object{BD+11\,3024}      & 52688-55093 & 357 & 0.033 \\
\hline\noalign{\smallskip}
  &  &  &  \\
     &  \multicolumn{3}{c}{ASAS-SN}   \\
Star &  HJD range & N & $\sigma$ \\
     & (24 00000+)&   &  (mag)   \\
\hline\noalign{\smallskip}
\object{BD+13\,5000}      & 56216-57932 & 354 & 0.029\\
\object{TYC\,3557-919-1}  & 56695-57950 & 371 & 0.009\\
\object{TYC\,5163-1764-1} & 56389-57931 & 387 & 0.009\\
\object{BD+11\,3024}      & 56337-57930 & 373 & 0.005\\
\hline\noalign{\smallskip}
  &  &  &  \\
     &  \multicolumn{3}{c}{T60}   \\
Star &  HJD range & N & $\sigma$ \\
     & (24 00000+)&   &  (mag)   \\
\hline\noalign{\smallskip}
\object{BD+13\,5000}      & 56506-57365 & 164 & 0.010\\
\object{TYC\,3557-919-1}  & 56536-57731 & 220 & 0.014\\
\object{TYC\,5163-1764-1} & 56533-57285 & 140 & 0.012\\
\object{BD+11\,3024}      & 56532-57534 & 107 & 0.007\\
\hline\noalign{\smallskip}
\end{tabular}
\end{center}
\end{table}

\begin{figure*}[!htb]
\centering
{\includegraphics[angle=0,scale=0.47,clip=true]{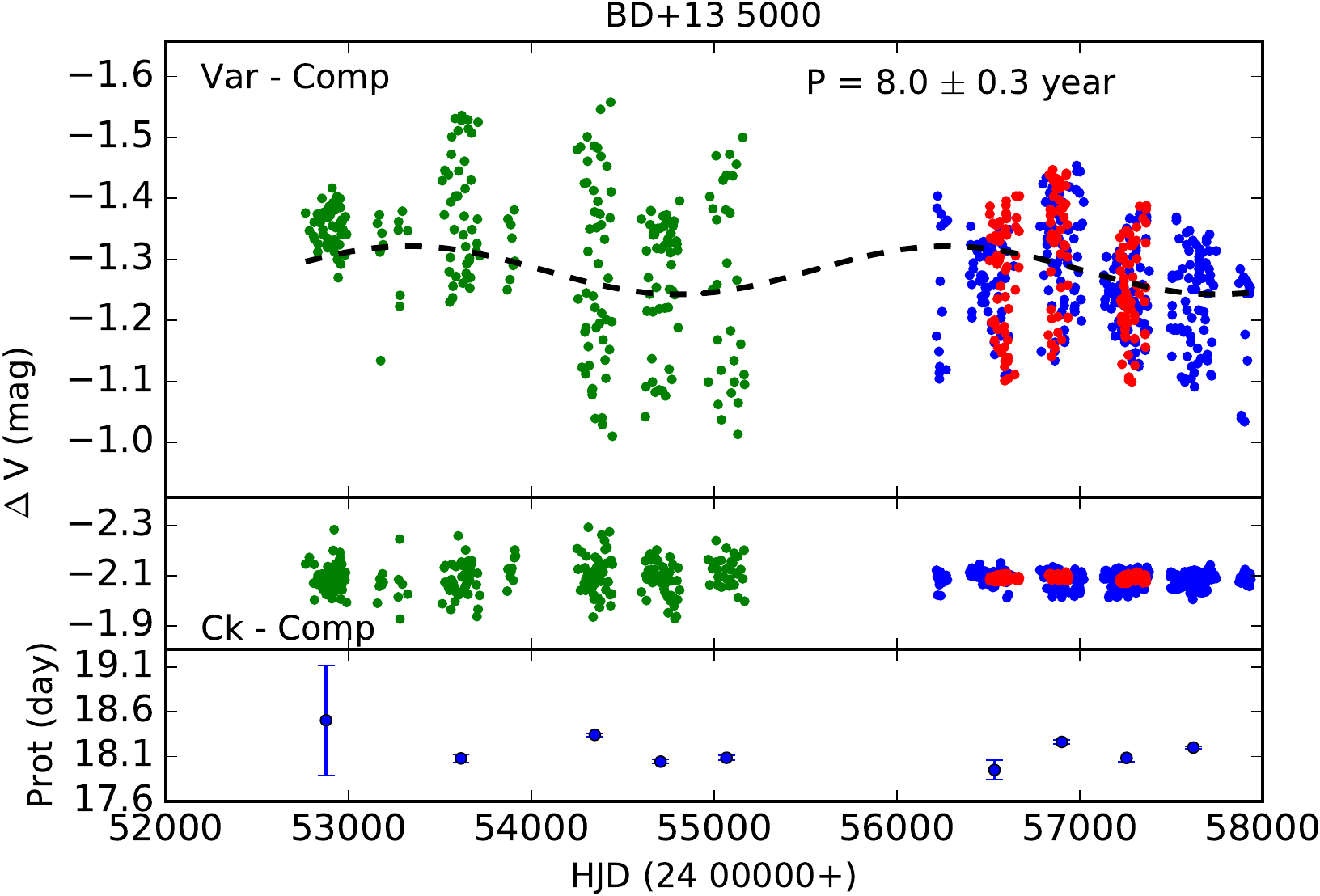}}
{\includegraphics[angle=0,scale=0.47,clip=true]{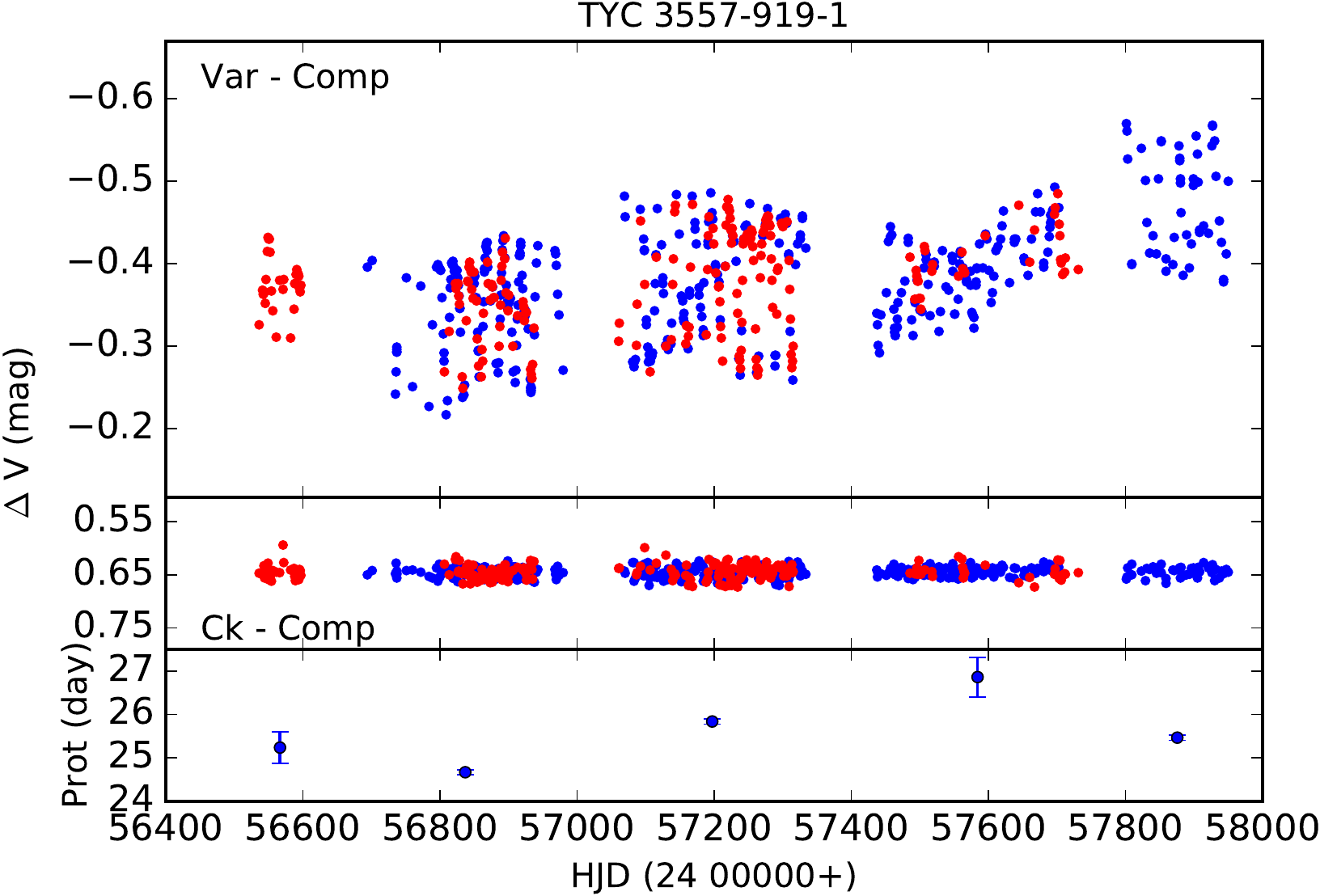}}\vspace{0.3cm}
{\includegraphics[angle=0,scale=0.47,clip=true]{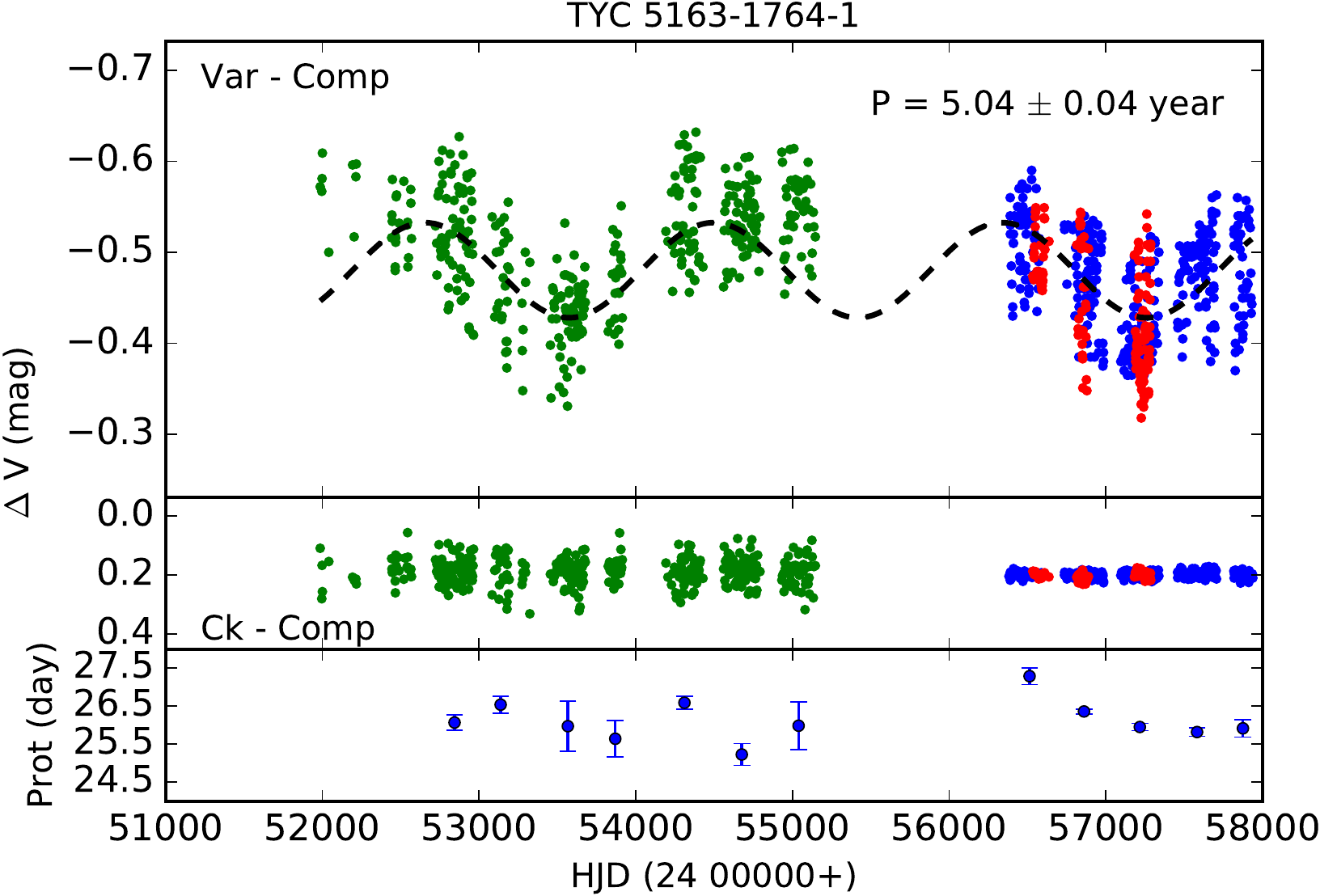}}
{\includegraphics[angle=0,scale=0.47,clip=true]{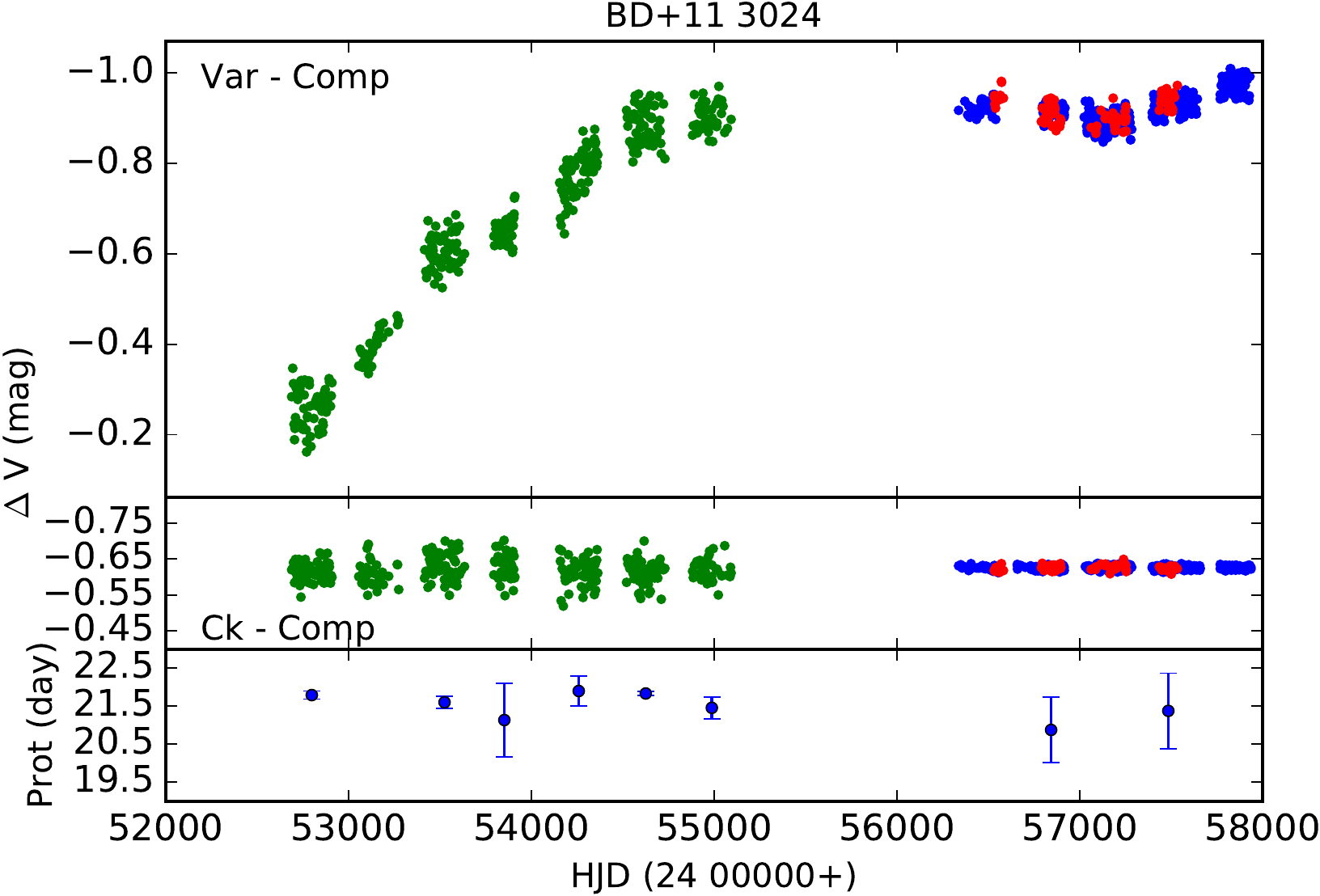}}
\caption{Differential photometry of program stars. Var, comp and ck denote variable, 
comparison and check stars, respectively. Black dashed lines inside plot windows 
of \object{BD+13\,5000} and \object{TYC\,5163-1764-1} indicate estimated period for 
long term brightness variation. The estimated periods are shown in each window together with 
their uncertainties. Green, red and blue filled points show photometry from ASAS, 
T60, and ASAS-SN, respectively. Lowermost panel of each window shows seasonal 
photometric periods (see Section 3.3.2, and Table~\ref{T5}).}\label{F3}
\end{figure*}

\subsection{T60 photometry}\label{S3.2}

We collect Bessell $V$ CCD observations with 0.6 m robotic telescope (T60) at TNO, equipped with 
2048 $\times$ 2048 pixels FLI ProLine 3041--UV CCD 
camera\footnote{\url{http://www.tug.tubitak.gov.tr/t60_fli_ccd.php}}
with a pixel size of 15 $\times$ 15 $\mu m^{2}$. We reduce automatically obtained CCD images with
$ccdred$ package in IRAF environment, following bias subtraction and then flat field division of 
science frames. The CCD has an efficient cooling system, which cools down the CCD to $-$60\degr~C 
and strictly stabilizes it during the whole night, hence bias and dark frame counts are almost 
same and we do not apply dark subtraction. We extract magnitudes of program stars 
from reduced science frames with $phot$ package in IRAF. Since comparison stars are just a few 
arc minutes away from target stars, we did not apply atmospheric extinction correction. 
In Table~\ref{T4}, we tabulate the brief information on our T60 observations.

We present T60, ASAS and ASAS-SN $V$ observations in Figure~\ref{F3}. Lower precision of ASAS 
measurements is noticeable compared to the T60 and ASAS-SN observations. Although the observational
scatter can be as large as amplitude of individual light curves for ASAS data, it still confirms 
global stability of comparison stars, but with lower precision. T60 and ASAS-SN data provide further 
confirmation for the stability of comparison stars, with better precision.

\subsection{Photometric analysis}\label{S3.3}

\subsubsection{Global photometric variation}\label{S3.3.1}

We observe variable mean brightness and light curve amplitude for all targets. \object{BD+13\,5000} and 
\object{TYC\,5163-1764-1} exhibit cyclic mean brightness variation in time. Application of 
Lomb-Scargle periodogram \citep{Lomb_1976Ap&SS, Scargle_1982ApJ} using $scipy$ package under 
$Python$ environment leads to 8.0$\pm$0.3 and 5.04$\pm$0.04 year cycle period for 
\object{BD+13\,5000} and \object{TYC\,5163-1764-1}, respectively. No evidence is found for mean 
brightness variation with a longer period. In the case of \object{TYC\,3557-919-1},
collected data cover only four years of time range, and variation pattern does not show any cyclic
variation, but a $\sim$0\fm1 global brightening trend.

\begin{figure}[!htb]
\centering
{\includegraphics[angle=0,scale=0.5,clip=true]{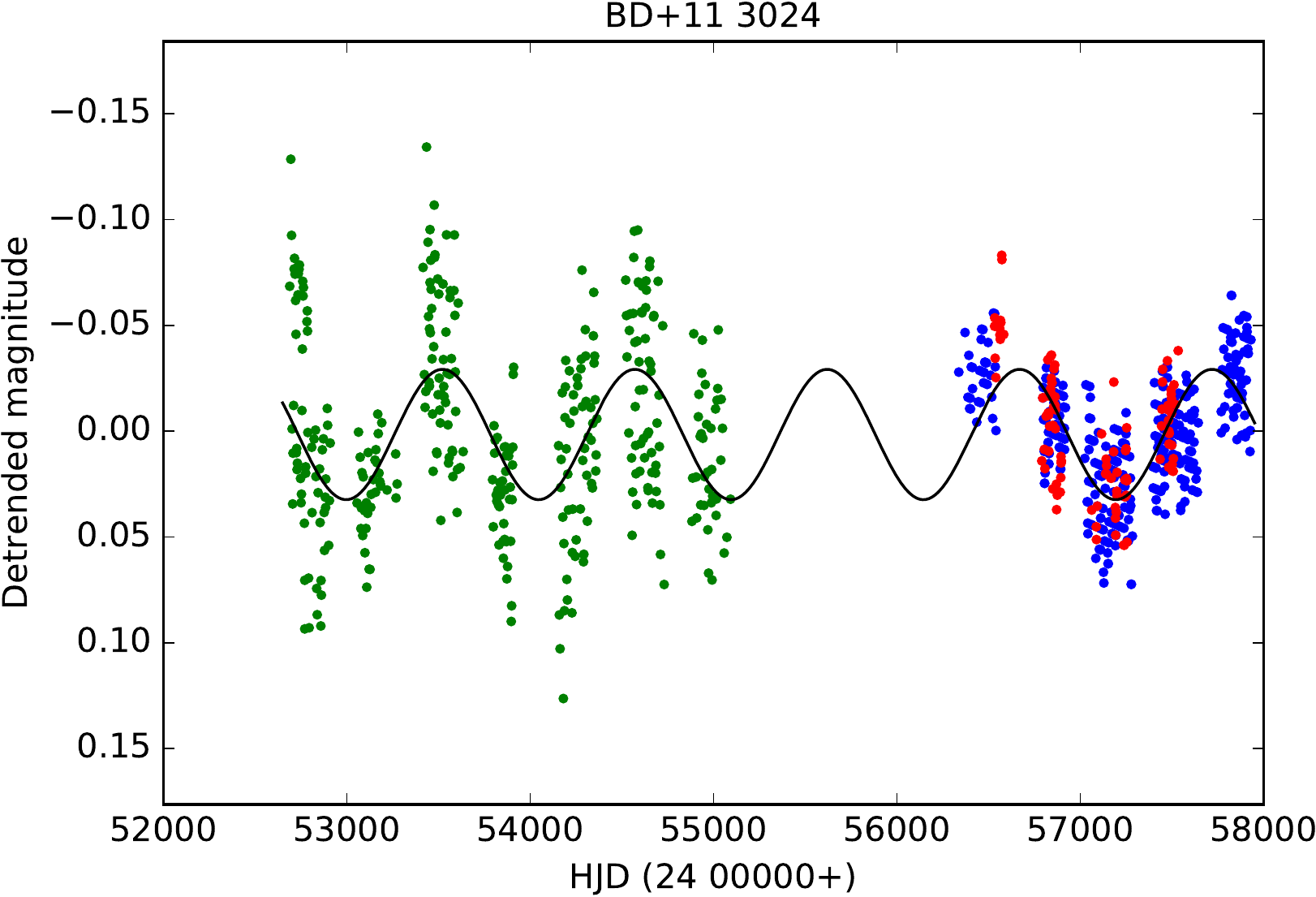}}
\caption{2.87-yr cyclic brightness variation (black curve) detected in the de-trended 
photometric data of \object{BD+11\,3024}. Meaning of colors are the same as Figure~\ref{F3}.}\label{F4}
\end{figure}

On the other hand, \object{BD+11\,3024} exhibits a striking increase in global brightness in 
ASAS photometry. The amount of increase is $\sim$0\fm7 and this increase is accompanied with a variable 
light curve amplitude from season to season. T60 and ASAS-SN observations show that the star somehow 
maintains its maximum brightness observed at the end of ASAS data. We apply second order polynomial 
fit to the ASAS data of the star, and linear fit to ASAS-SN and T60 data, to de-trend the whole light 
curve, and search for a signal of any cyclic brightness variation. Application of Lomb-Scargle periodogram 
to the de-trended data indicates $2.87\pm0.12$ year cyclic variation with an amplitude of 0\fm03, 
which is shown in Figure~\ref{F4}.

\begin{table*}
\caption{Brightness levels and photometric periods found from application of the ANOVA method to the 
seasonal light curves. In the first four columns, set number, beginning, end and mean heliocentric 
Julian dates of the corresponding light curve are given. Photometric period and its uncertainty 
($P$ and $\sigma~(P)$) are in unit of day. The maximum, minimum and mean brightnesses, and the light 
curve amplitude ($A$) are given in magnitudes. In the last column, $N$ shows number of 
data points used in the analysis.}\label{T5}
\begin{center}
\begin{tabular}{ccccccccccc}
\hline\noalign{\smallskip}
& & & & & & & & & & \\
\multicolumn{11}{l}{\object{BD+11\,3024}} \\
Set   &   HJD min.   &   HJD max.   &   HJD mean   &   $P$   &   $\sigma~(P)$   &   max   &   min   &   mean   &   $A$   &   $N$   \\
& (24 00000+) & (24 00000+) & (24 00000+) & & & & & & & \\
\hline\noalign{\smallskip}
1   &   52688.8799   &   52909.4910   &   52799.1855   &   21.79   &   0.11   &   -0.323   &   -0.206   &   -0.265   &   0.117   &   65   \\
2   &   53415.8809   &   53634.4802   &   53525.1806   &   21.60   &   0.16   &   -0.650   &   -0.553   &   -0.602   &   0.097   &   62   \\
3   &   53794.8888   &   53909.7268   &   53852.3078   &   21.14   &   0.97   &   -0.677   &   -0.622   &   -0.649   &   0.055   &   39   \\
4   &   54154.8879   &   54363.4964   &   54259.1922   &   21.90   &   0.39   &   -0.825   &   -0.731   &   -0.778   &   0.094   &   57   \\
5   &   54520.8874   &   54730.4803   &   54625.6839   &   21.84   &   0.05   &   -0.938   &   -0.831   &   -0.884   &   0.107   &   60   \\
6   &   54881.8918   &   55092.4985   &   54987.1952   &   21.46   &   0.29   &   -0.952   &   -0.874   &   -0.913   &   0.078   &   38   \\
7   &   56789.5162   &   56896.2329   &   56842.8746   &   20.88   &   0.86   &   -0.926   &   -0.875   &   -0.901   &   0.051   &   37   \\
8   &   57434.6040   &   57533.3902   &   57483.9971   &   21.38   &   0.99   &   -0.964   &   -0.923   &   -0.943   &   0.041   &   27   \\
\hline\noalign{\smallskip}
& & & & & & & & & & \\
\multicolumn{11}{l}{\object{BD+13\,5000}} \\
Set   &   HJD min.   &   HJD max.   &   HJD mean   &   $P$   &   $\sigma~(P)$   &   max   &   min   &   mean   &   $A$   &   $N$   \\
& (24 00000+) & (24 00000+) & (24 00000+) & & & & & & & \\
\hline\noalign{\smallskip}
1   &   52764.9261   &   52989.5301   &   52877.2281   &   18.51   &   0.61   &   -1.378   &   -1.320   &   -1.349   &   0.058   &   52   \\
2   &   53512.9205   &   53717.5263   &   53615.2234   &   18.08   &   0.04   &   -1.507   &   -1.263   &   -1.385   &   0.244   &   45   \\
3   &   54250.9308   &   54443.5358   &   54347.2333   &   18.34   &   0.02   &   -1.506   &   -1.062   &   -1.284   &   0.444   &   52   \\
4   &   54601.9120   &   54811.5315   &   54706.7218   &   18.04   &   0.02   &   -1.360   &   -1.047   &   -1.203   &   0.313   &   50   \\
5   &   54967.9262   &   55166.5666   &   55067.2464   &   18.09   &   0.03   &   -1.451   &   -1.063   &   -1.257   &   0.388   &   31   \\
6   &   56397.1295   &   56669.1541   &   56533.1418   &   17.95   &   0.11   &   -1.352   &   -1.160   &   -1.256   &   0.192   &   110   \\
7   &   56782.0933   &   57020.6935   &   56901.3934   &   18.26   &   0.03   &   -1.424   &   -1.171   &   -1.298   &   0.253   &   132   \\
8   &   57132.1243   &   57378.7314   &   57255.4279   &   18.09   &   0.04   &   -1.349   &   -1.159   &   -1.254   &   0.190   &   161   \\
9   &   57496.1318   &   57745.6977   &   57620.9148   &   18.20   &   0.02   &   -1.337   &   -1.094   &   -1.215   &   0.243   &   83   \\
\hline\noalign{\smallskip}
& & & & & & & & & & \\                                                    
\multicolumn{11}{l}{\object{TYC\,3557-919-1}} \\
Set   &   HJD min.   &   HJD max.   &   HJD mean   &   $P$   &   $\sigma~(P)$   &   max   &   min   &   mean   &   $A$   &   $N$   \\
& (24 00000+) & (24 00000+) & (24 00000+) & & & & & & & \\
\hline\noalign{\smallskip}
1   &   56536.2497   &   56597.1793   &   56566.7145   &   25.24   &   0.36   &   -0.431   &   -0.292   &   -0.361   &   0.139   &   29   \\
2   &   56694.1694   &   56979.6930   &   56836.9312   &   24.67   &   0.05   &   -0.412   &   -0.251   &   -0.332   &   0.161   &   179   \\
3   &   57060.6618   &   57333.7025   &   57197.1822   &   25.84   &   0.06   &   -0.462   &   -0.283   &   -0.373   &   0.179   &   203   \\
4   &   57437.1658   &   57731.1923   &   57584.1791   &   26.86   &   0.46   &   -0.432   &   -0.363   &   -0.398   &   0.069   &   132   \\
5   &   57801.1689   &   57949.8032   &   57875.4860   &   25.46   &   0.06   &   -0.567   &   -0.398   &   -0.483   &   0.169   &   48   \\
\hline\noalign{\smallskip}                                                      
& & & & & & & & & & \\                                                    
\multicolumn{11}{l}{\object{TYC\,5163-1764-1}} \\
Set   &   HJD min.   &   HJD max.   &   HJD mean   &   $P$   &   $\sigma~(P)$   &   max   &   min   &   mean   &   $A$   &   $N$   \\
& (24 00000+) & (24 00000+) & (24 00000+) & & & & & & & \\
\hline\noalign{\smallskip}
1   &   52721.8999   &   52964.5131   &   52843.2065   &   26.07   &   0.20   &   -0.594   &   -0.463   &   -0.528   &   0.131   &   66   \\
2   &   53082.9093   &   53191.7672   &   53137.3383   &   26.54   &   0.22   &   -0.541   &   -0.392   &   -0.467   &   0.149   &   25   \\
3   &   53454.9023   &   53677.5511   &   53566.2267   &   25.98   &   0.65   &   -0.457   &   -0.382   &   -0.420   &   0.075   &   61   \\
4   &   53823.8992   &   53913.7795   &   53868.8394   &   25.64   &   0.48   &   -0.535   &   -0.407   &   -0.471   &   0.128   &   22   \\
5   &   54191.8947   &   54428.5135   &   54310.2041   &   26.59   &   0.18   &   -0.604   &   -0.488   &   -0.546   &   0.116   &   53   \\
6   &   54559.9012   &   54792.5118   &   54676.2065   &   25.23   &   0.29   &   -0.574   &   -0.494   &   -0.534   &   0.080   &   63   \\
7   &   54931.8943   &   55145.5269   &   55038.7106   &   25.99   &   0.62   &   -0.593   &   -0.500   &   -0.546   &   0.093   &   42   \\
8   &   56389.0940   &   56636.1713   &   56512.6327   &   27.29   &   0.22   &   -0.553   &   -0.461   &   -0.507   &   0.092   &   102   \\
9   &   56735.1287   &   56984.6854   &   56859.9071   &   26.36   &   0.07   &   -0.526   &   -0.376   &   -0.451   &   0.150   &   118   \\
10   &   57094.9021   &   57337.5197   &   57216.2109   &   25.95   &   0.09   &   -0.501   &   -0.364   &   -0.433   &   0.137   &   170   \\
11   &   57457.1418   &   57705.5100   &   57581.3259   &   25.82   &   0.11   &   -0.525   &   -0.406   &   -0.465   &   0.119   &   91   \\
12   &   57817.1547   &   57930.9784   &   57874.0666   &   25.92   &   0.23   &   -0.554   &   -0.400   &   -0.477   &   0.154   &   46   \\
\noalign{\smallskip}\hline
\end{tabular}
\end{center}
\end{table*}

\subsubsection{Photometric period and seasonal brightness variations}\label{S3.3.2}

For each target, we determine minimum, maximum and mean brightnesses of seasonal light curves, 
as well as corresponding light curve amplitudes and photometric periods. In most cases, light curves
are asymmetric, which is a common feature observed in light curves of chromospherically 
active stars. Hence, application of periodogram methods 
based on Fourier transform would result in multiple periods, i.e. main period, which corresponds to the
photometric period, and its harmonics. However, application of statistical methods, such as 
Phase Dispersion Minimization \citep[PDM,][]{PDM_Stellingwerf_1978ApJ} and Analysis of Variances 
\citep[ANOVA,][]{ANOVA_Czerny_1996ApJ}, are more efficient to find a single period which would be
the single representative period for the observed variation, without any additional periods 
(i.e. harmonics). In the scope of our study, we adopt ANOVA method. In the first step of the analysis, 
we determine photometric period of each observing season. Second, we phased the light curve of that 
season with its resulting photometric period. Finally, we fit a cubic spline to the phase-folded 
light curve, and determine minimum and maximum brightnesses. The mean brightness and the amplitude 
of the light curve are calculated from determined minimum and maximum brightnesses. We tabulate our 
analysis results in Table~\ref{T5}. We plot seasonal light curves of each target in the appendix.

For any of the target star, the number of sets listed in Table~\ref{T5} is insufficient 
for evaluation of any correlation between parameters, e.g. the photometric period and 
the light curve amplitude, the amplitude and the brightness, or the photometric period
and the mean brightness. For example, plotted photometric periods in Figure~\ref{F3} do not 
seem to be correlated with the mean brightness. However, considerable discontinuities 
in the photometric data, and observing seasons where the photometric period could not be 
determined due to the very low light curve amplitude reduce the number of useful datasets 
for analysis, thus may cause misinterpretation of the real situation. Still, we find average 
photometric periods as 21\fd49, 18\fd17, 25\fd61 and 26\fd11 for \object{BD+11\,3024}, 
\object{BD+13\,5000}, \object{TYC\,3557-919-1} and \object{TYC\,5163-1764-1}, 
respectively. We use the distribution range of photometric periods and estimate 
$\Delta P/P=(P_{max} - P_{min})/P_{eq}$, where $P_{max}$ and $P_{min}$ denote calculated
maximum and minimum photometric periods for corresponding target, and $P_{eq}$ is equatorial
rotation period, which is adopted as the mean value of the measured rotation periods for each 
target, given in Table~\ref{T5}. Using periods listed in Table~\ref{T5}, we find $\Delta P/P$ 
values as 0.047, 0.031, 0.086, 0.079 for \object{BD+11\,3024}, \object{BD+13\,5000}, 
\object{TYC\,3557-919-1} and \object{TYC\,5163-1764-1}, respectively.

\section{Summary and discussion}\label{S4}

We present analysis of spectroscopic and long-term photometric observations of four cool stars,
\object{BD+11\,3024}, \object{BD+13\,5000}, \object{TYC\,3557-919-1} and 
\object{TYC\,5163-1764-1}. Strong emission in H$_{\alpha}$ and 
Ca\,{\sc ii} H\& K lines, which exceeds the continuum in the case of \object{BD+13\,5000} 
and \object{TYC\,3557-919-1}, and remarkable changes in mean brightness and light curve 
amplitude clearly indicate chromospheric activity on these stars. Observed emissions
appear to be rotationally modulated. Furthermore, non-stable radial velocities of the 
target stars indicate that they might possibly be a member of a binary system. However, 
binary nature of these stars and rotational modulation of the emission features need 
to be confirmed by further observations, which would cover at least a rotational cycle.

Analysis of seasonal light curves makes possible to trace seasonal photometric periods. 
Distribution range of photometric periods enables us to calculate $\Delta P/P$, which 
gives some hints on the lower limit of the differential rotation, and spot evolution 
in time, which could also alter the distribution of the photometric period in time.
Growth and decay of spot regions at different latitudes and longitudes (i.e. temporal 
and spatial evolution of spots) may cause variation in photometric period 
\citep[see, e.g.][]{FG_IS_Fekel_et_al_2002AJ}.
Less number of datasets in Table~\ref{T5} do not allow one to arrive at any conclusive 
results, however, we notice that calculated $\Delta P/P$ values are larger than the 
values of spotted stars studied in \citet{Hall_Busby_1990_difrot}.

We find average rotational period of \object{BD+11\,3024} as 21\fd49. Considering the 
uncertainties of seasonal periods in the Table~\ref{T5}, it is in agreement with the 
period reported by \citet{gsc3557_gsc969_Bernhard_2008OEJV}. We observe more precise
agreement in the case of \object{BD+13\,5000}, where our average period is 18\fd17 and the 
period reported by \citet{Kiraga_ASAS_ROSAT_sources_2012AcA} is 18\fd14. The average
photometric period of \object{TYC\,3557-919-1} found in this study is 25\fd61, while
it was found as 25\fd08 by \citet{gsc3557_gsc969_Bernhard_2008OEJV}. In summary, we observe
no discrepancies between average photometric periods found in this study and the literature
values.

Below, we discus our further findings for each star, in a separate paragraph.

\emph{\object{BD+13\,5000}.} Our atmospheric analysis indicates K0IV spectral type for
the star. Measured radial velocities are distributed in a range of 90 \kms\@, indicating
that the star could be a spectroscopic binary. Very strong emission, which exceeds 
the continuum, is observed in H$_{\alpha}$ and Ca\,{\sc ii} H\& K lines. These features 
are accompanied with a cyclic variation in the mean brightness with a period of 
$8.0\pm0.3$ year, which repeats itself for two times in the time base of the current 
photometric data, and indicates an activity cycle. In most cases, light curve amplitudes 
are larger than 0\fm19. All these findings suggest a very strong and cyclic chromospheric 
activity.

\emph{\object{TYC\,3557-919-1}.} Analysis of TFOSC spectra indicates K0III-IV spectral 
type. Radial velocities appear to be variable, where the maximum difference among the
measurements is to be 52 \kms\@, suggesting the possibility of being spectroscopic binary. 
Similar to \object{BD+13\,5000}, H$_{\alpha}$ and Ca\,{\sc ii} H\& K lines exhibit very 
strong emission that exceeds the continuum. Available photometric data cover four years 
and only indicate a global brightening trend of $\sim$0\fm1, with a seasonal light curve 
amplitude usually between 0\fm1 and 0\fm2. These are clear evidence of very strong 
chromospheric activity.

\emph{\object{TYC\,5163-1764-1}.} Spectral type of the star is found as K2III. As in the
cases of \object{BD+13\,5000} and \object{TYC\,3557-919-1}, we observe unstable radial 
velocities in a range of 49 \kms\@, indicating possibility of being spectroscopic binary.
Emission observed in H$_{\alpha}$ and Ca\,{\sc ii} H\& K lines are weaker compared to
\object{BD+13\,5000} and \object{TYC\,3557-919-1}, but still clearly observed, 
reaching to the continuum level around Ca\,{\sc ii} K line. We find cyclic mean 
brightness variation with a period of $5.04\pm0.04$ year, repeating itself more than 
three times in the time span of the photometric data. Observed light curve amplitudes 
are between 0\fm075 and 0\fm154. These findings suggest strong and cyclic
chromospheric activity.

\emph{\object{BD+11\,3024}.} TFOSC spectra of this star indicate K0III-IV spectral type.
Measured radial velocities are distributed between smaller range (17 \kms\@) compared to 
the other target stars, weakly indicates possibility of being spectroscopic binary.
Strengths of H$_{\alpha}$ and Ca\,{\sc ii} H\& K emissions of the star are very similar 
to the observed emissions in \object{TYC\,5163-1764-1}. Photometric data of the star 
exhibit very dramatic brightness increase ($\sim$0\fm7) accompanied with a $2.87\pm0.12$ year
cyclic brightness variation with 0\fm03 amplitude. The star might have multiple cycles
if the global brightness increase is part of a longer term cyclic brightness change with
a larger amplitude. This would not be surprising since multiple cycles appear to be observed 
often in active stars \citep{Olah_cycles_2009A&A}. Seasonal light curve amplitude, usually 
below 0\fm1, and interesting brightness variations make this star an attractive target for 
further studies.

Spectroscopic and photometric characteristics of \object{BD+13\,5000}, 
\object{TYC\,3557-919-1} and \object{TYC\,5163-1764-1} are similar to the typical RS\,CVn 
stars with strong emission in their Ca\,{\sc ii} H\& K lines, cyclic mean brightness variation
and remarkable light curve amplitude, e.g. V1149\,Ori \citep{V1149Ori_Fekel_et_al_2005AJ, 
Jetsu_Henry_long_term_data_2017ApJ}, HD\,208472 \citep{Ozdarcan_V2075Cyg_2010AN, 
DI_V2075Cyg_Ozdarcan2016}, FG\,UMa \citep{FG_IS_Fekel_et_al_2002AJ, Ozdarcan_FGUMa_2012AN}, 
and many other targets studied in \citet{Olah_cycles_2009A&A} and
\citet{Jetsu_Henry_long_term_data_2017ApJ}. Individually, photometric behaviour of 
\object{BD+11\,3024} resembles RS\,CVn binary IS\,Vir \citep{FG_IS_Fekel_et_al_2002AJ,
ISVir_Olah_et_al_2013AN}, which possesses similar light curve amplitude, and a long term 
variation with an estimated period of 5-6 years \citep{ISVir_Olah_et_al_2013AN}. Dramatic 
changes in the mean brightness of \object{BD+11\,3024} suggests strong spot activity. In this 
case, low light curve amplitude could be explained by either assuming that the distribution 
of spots is usually homogeneous on the surface of the star, or assuming the inclination 
of the rotational axis is low, as in the case of IS\,Vir, thus causing low light curve 
amplitude.

High resolution optical spectroscopy is required to determine more precise atmospheric 
properties and projected rotational velocities, as well as to trace radial velocities in
order to check whether each of these stars is single or a member of a binary system. 
Determination of projected rotational velocities from high resolution spectroscopy 
would help to restrict the stellar radius, which is one of the key parameter to 
calculate stellar luminosity, i.e. vertical position of a star in HR diagram.
Further time series photometric observations would enable one to trace activity cycles
and study the relation between cycle length and rotation period 
\citep[e.g.][]{Olah_strassmeier_cycles_2002AN}, as well as surface differential rotation
via distribution of seasonal photometric periods in time \citep{Hall_Busby_1990_difrot, 
Ozdarcan_V2075Cyg_2010AN, Ozdarcan_FGUMa_2012AN}.

\acknowledgements
We thank to T\"UB\.ITAK for a partial support in using RTT150 (Russian-Turkish 1.5-m telescope 
in Antalya) with project number 14BRTT150-678, and T60 telescope with project number 13CT60-504.
We also thank our referee, Dr. Katalin Ol{\'a}h, for her useful comments that improve the paper. 
This research has made use of the SIMBAD database, operated at CDS, Strasbourg, France.


\bibliographystyle{an}
\bibliography{rscvn_ok}

%
%



\appendix

\section{Seasonal light curves and phase folded radial velocities}\label{Ap}

\begin{figure}[!htb]
\centering
{\includegraphics[angle=0,scale=0.45,clip=true]{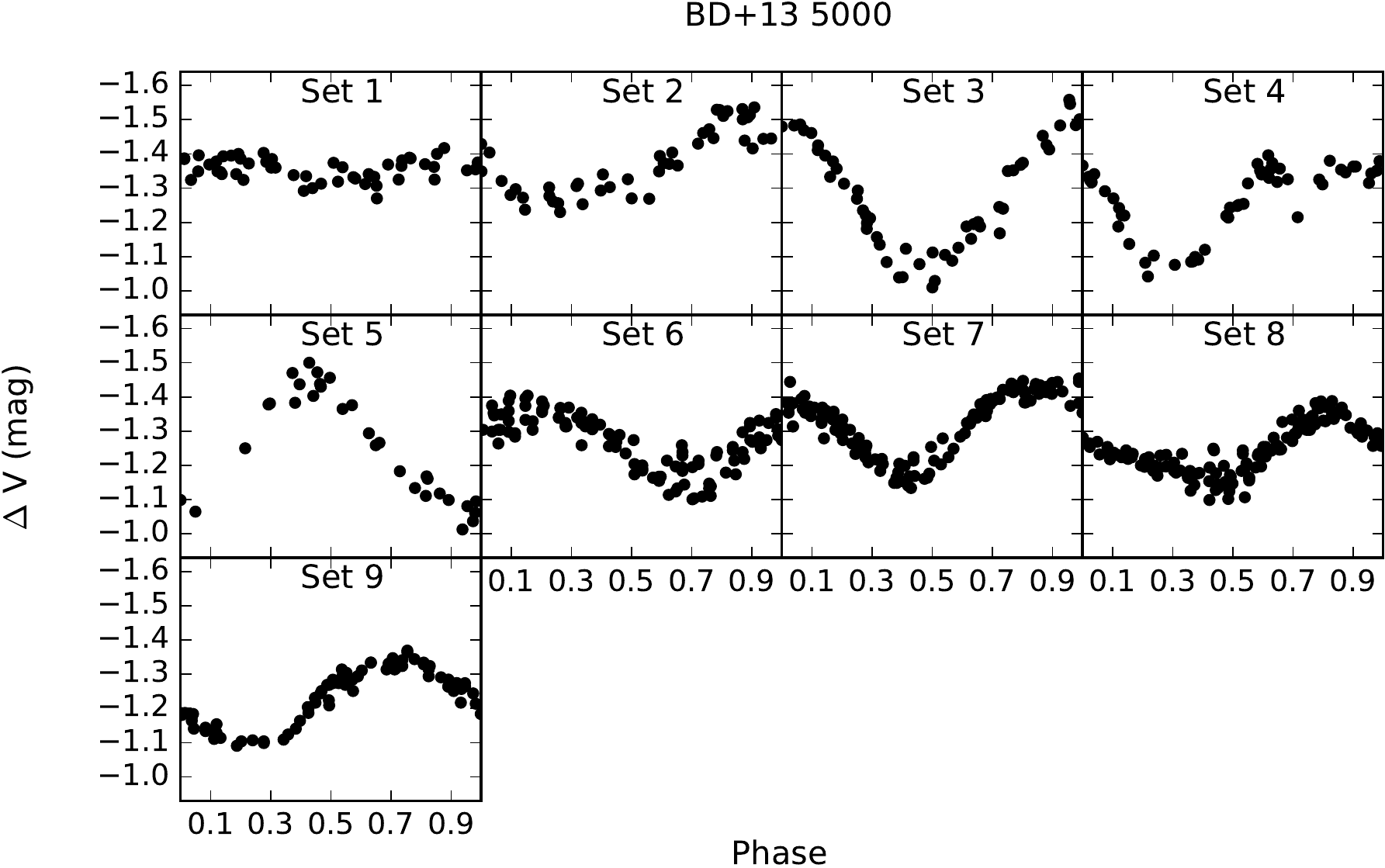}}\vspace{0.5cm}
{\includegraphics[angle=0,scale=0.45,clip=true]{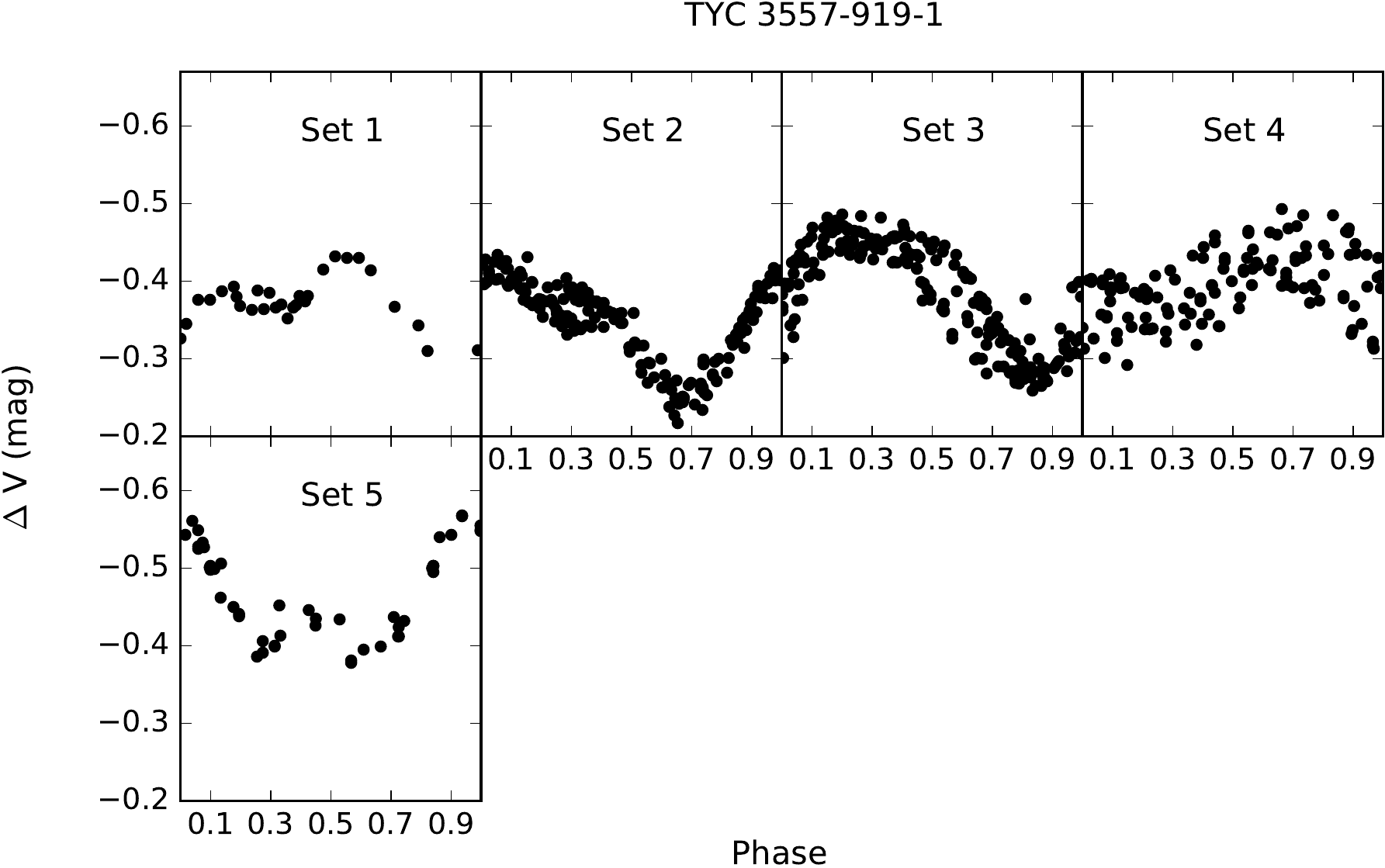}}\vspace{0.5cm}
{\includegraphics[angle=0,scale=0.45,clip=true]{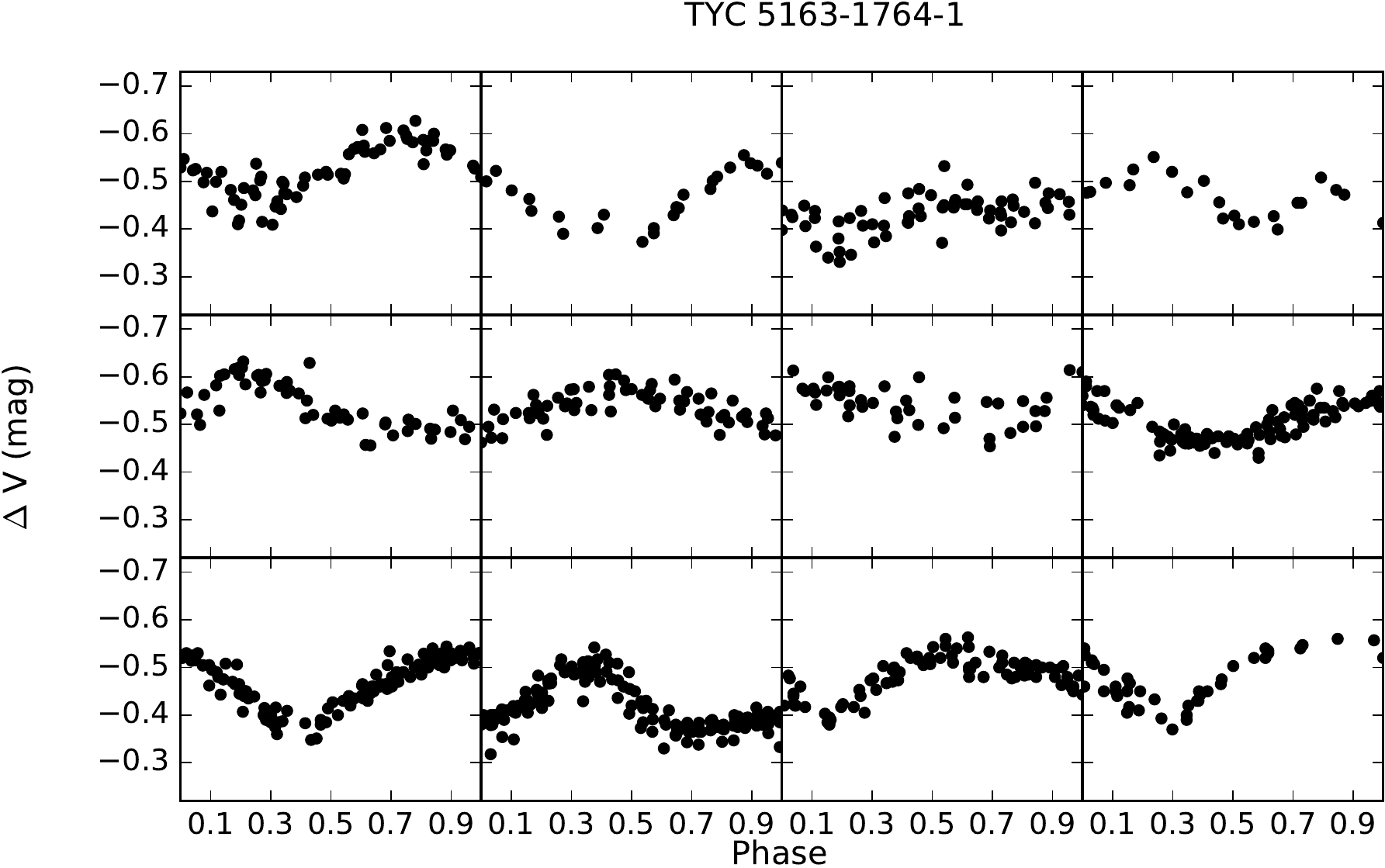}}\vspace{0.5cm}
{\includegraphics[angle=0,scale=0.45,clip=true]{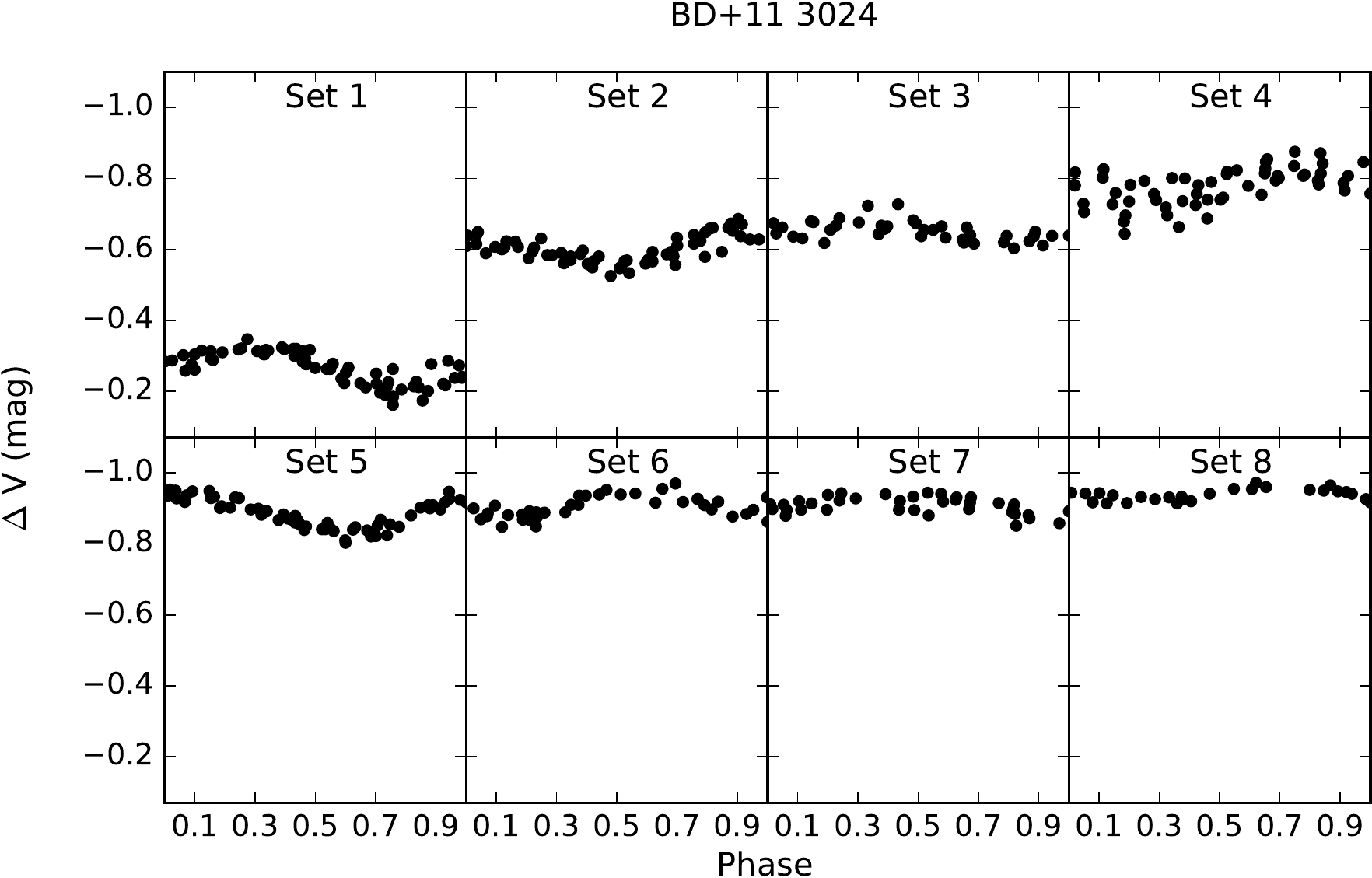}}
\caption{Phase folded seasonal light curves of the target stars. Phase folding
is done with the calculated photometric period listed in Table~\ref{T5} and
heliocentric Julian date of the initial data point of corresponding light curve.}\label{F_AP}
\end{figure}

\begin{figure}[!htb]
\centering
{\includegraphics[angle=0,scale=0.45,clip=true]{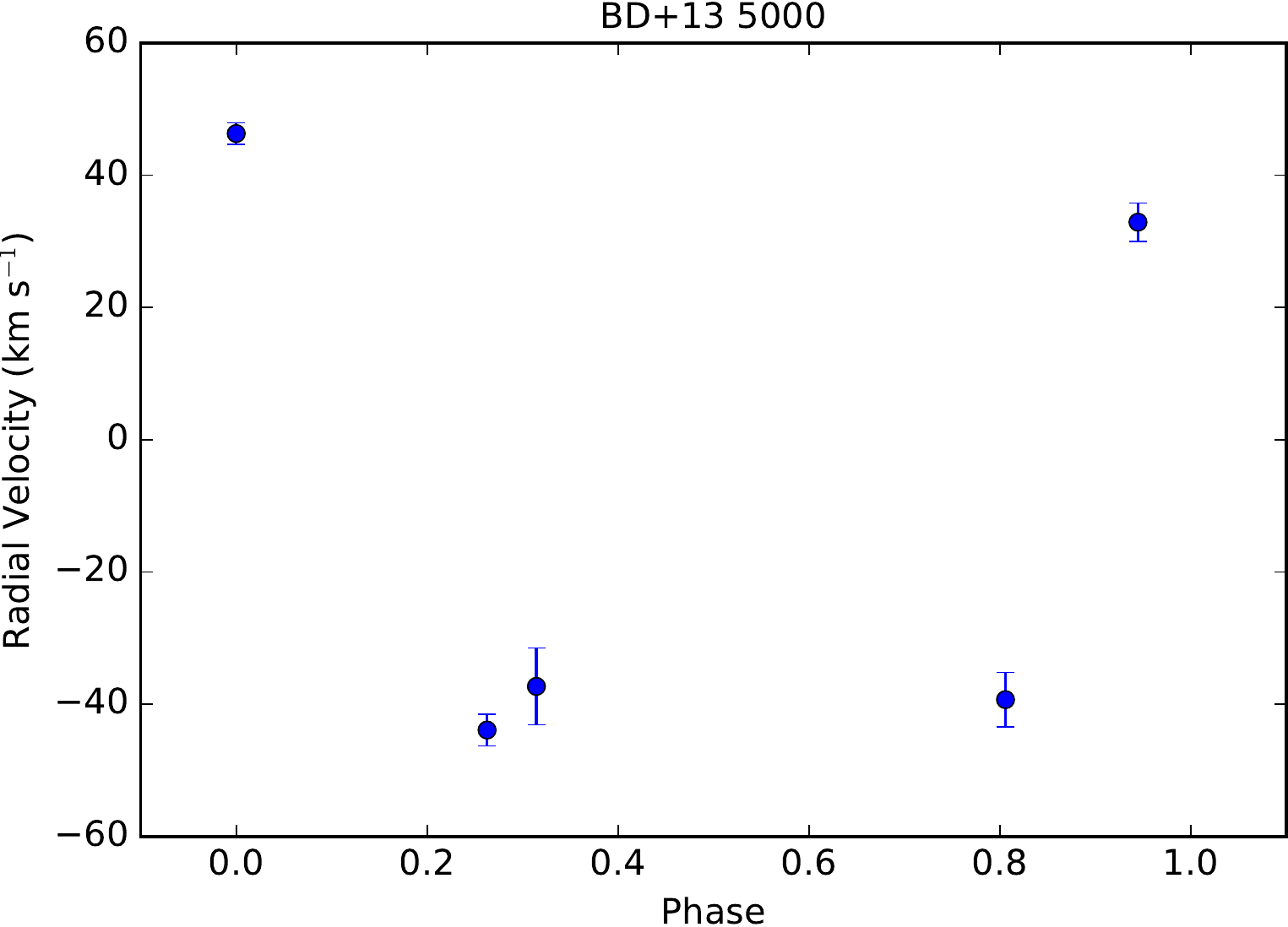}}\vspace{0.5cm}
{\includegraphics[angle=0,scale=0.45,clip=true]{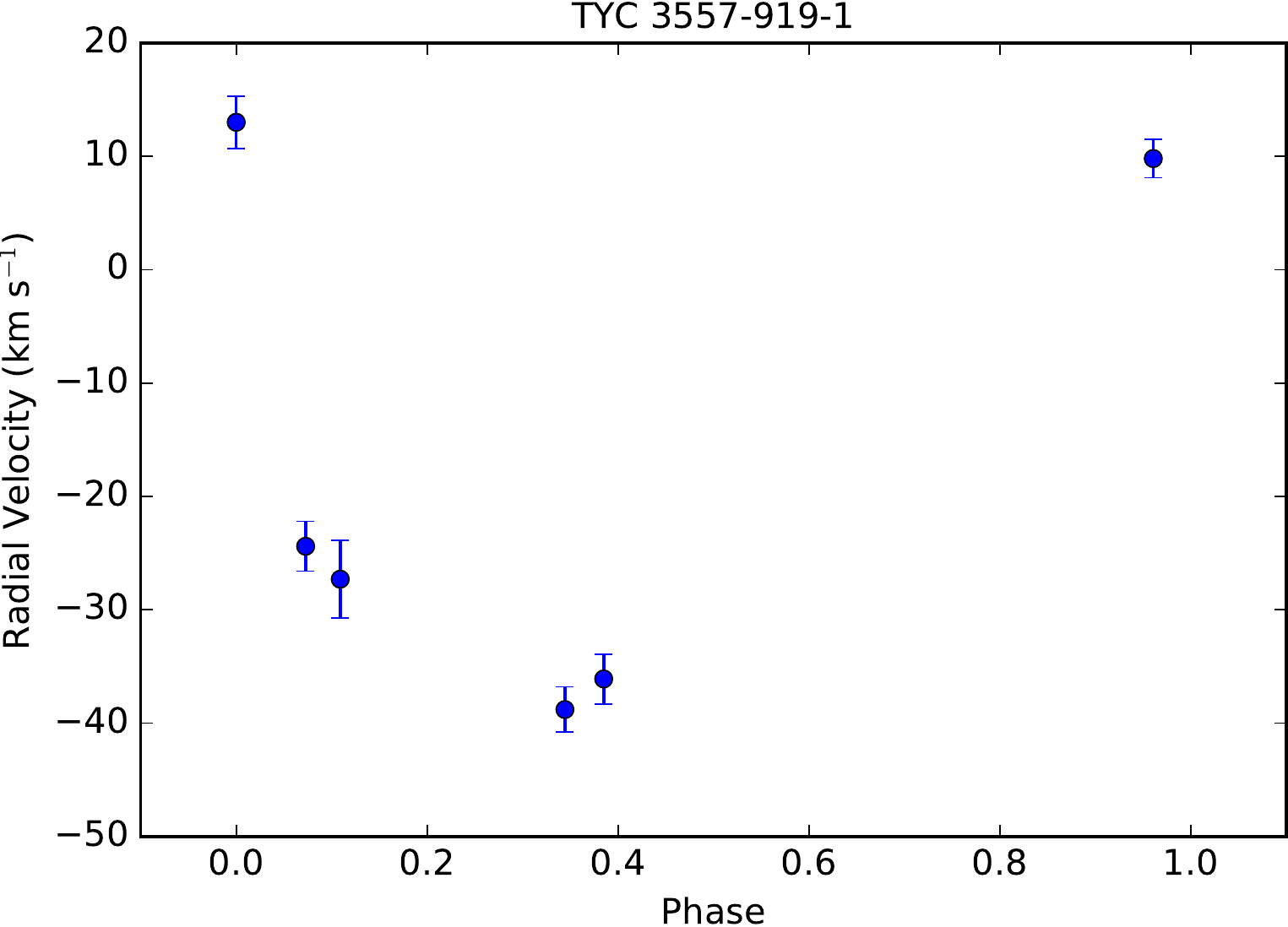}}\vspace{0.5cm}
{\includegraphics[angle=0,scale=0.45,clip=true]{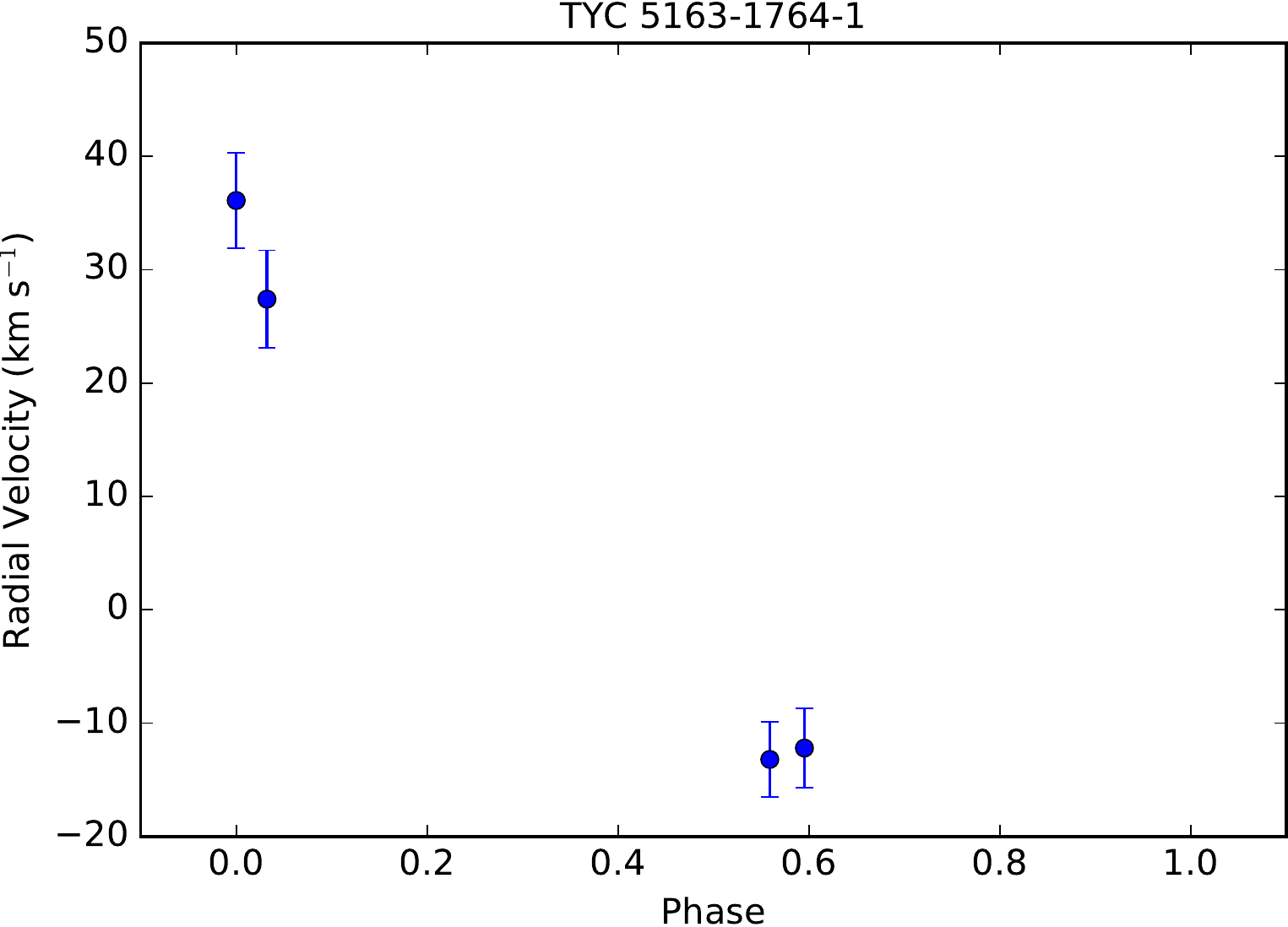}}\vspace{0.5cm}
{\includegraphics[angle=0,scale=0.45,clip=true]{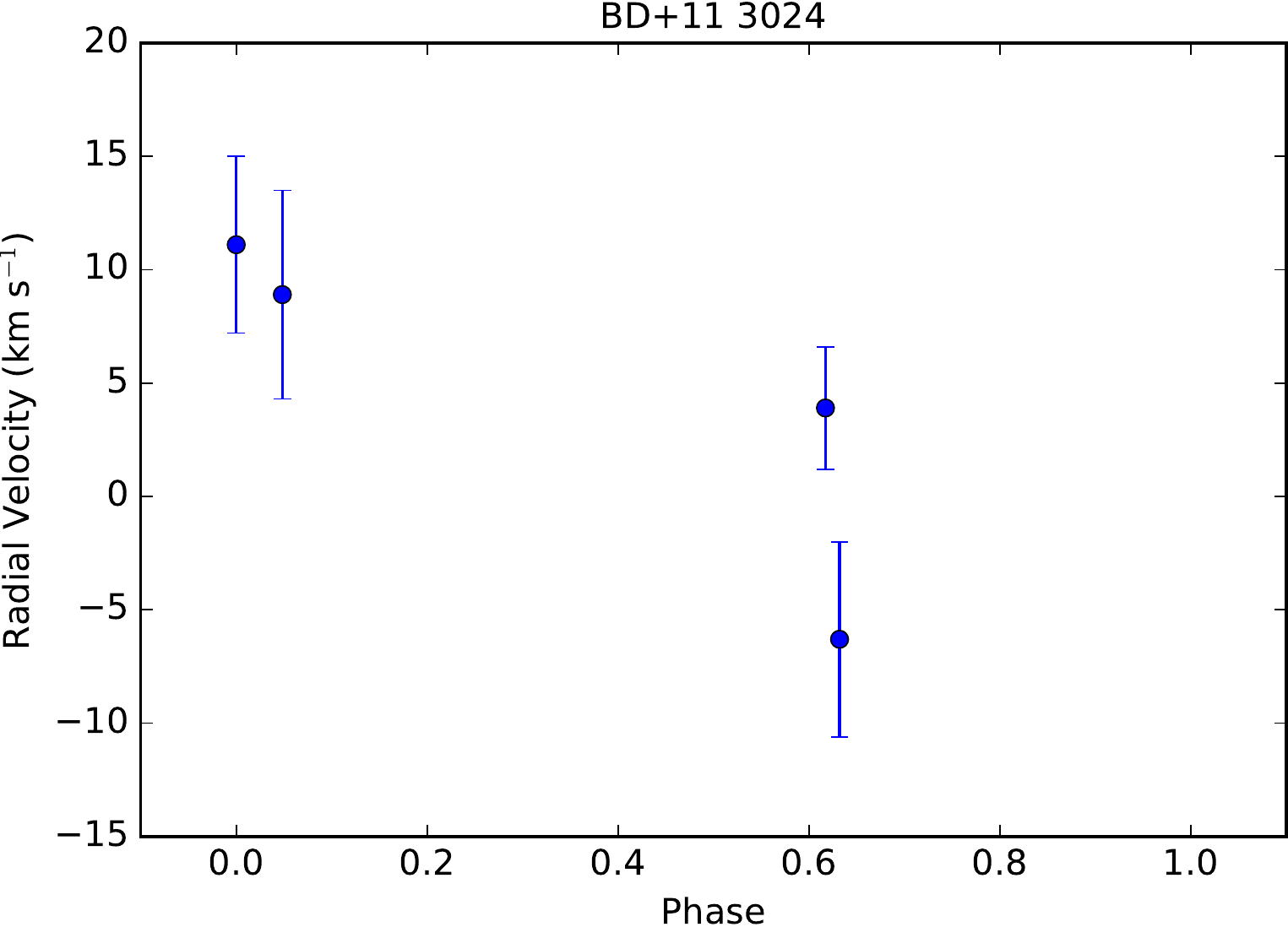}}
\caption{Phase folded radial velocities of the target stars. Phase folding is done
with respect to the time of the observed positive maximum velocity 
(ephemeris, see Table~\ref{T1}) and calculated mean photometric period read 
from Table~\ref{T5}.}\label{F_AP2}
\end{figure}

\end{document}